\documentclass[aps,prx,superscriptaddress,twocolumn]{revtex4-2}
\usepackage{graphicx,color}
\usepackage{verbatim}
\usepackage{amssymb}   
\usepackage{amsmath}
\usepackage{amsfonts}
\usepackage{mathdots}
\usepackage{hyperref}
\usepackage{multirow}
\usepackage{epsfig}
\usepackage{hyperref}
\usepackage{xcolor}
\usepackage{cancel}
\usepackage{booktabs}
\usepackage{array}
\usepackage{pifont}
\newcommand{\xmark}{\ding{55}}%
\usepackage{array, makecell} 

\usepackage{multirow}

\usepackage{braket}
\usepackage{bm}

\usepackage{multirow}

\begin{document}
	\title{Photon-drag photovoltaic effects and quantum geometric nature}

	\author{Ying-Ming Xie}\thanks{yingming.xie@riken.jp}
	\affiliation{RIKEN Center for Emergent Matter Science (CEMS), Wako, Saitama 351-0198, Japan} 	
	\author{Naoto Nagaosa}\thanks{nagaosa@riken.jp}
	\affiliation{RIKEN Center for Emergent Matter Science (CEMS), Wako, Saitama 351-0198, Japan} 
	\affiliation{Fundamental Quantum Science Program, TRIP Headquarters, RIKEN, Wako 351-0198, Japan} 	
	
	\date{\today}
	\begin{abstract}
		The bulk photovoltaic effect (BPVE) generates a direct current photocurrent under uniform irradiation and is a nonlinear optical effect traditionally studied in non-centrosymmetric materials. The two main origins of BPVE are the shift and injection currents, arising from transitions in electron position and electron velocity during optical excitation, respectively. Recently, it was proposed that photon-drag effects could unlock BPVE in centrosymmetric materials. However, experimental progress remains limited. In this work, we provide a comprehensive theoretical analysis of photon-drag effects inducing  BPVE (photon-drag BPVE). Notably, we find that photon-drag BPVE can be directly linked to quantum geometric tensors. Additionally, we propose that photon-drag shift currents can be fully isolated from other current contributions in non-magnetic centrosymmetric materials. We apply our theory explicitly to the 2D topological insulator $1T'$-WTe$_2$. Furthermore, we investigate photon-drag BPVE in a centrosymmetric magnetic Weyl semimetal, where we demonstrate that linearly polarized light generates photon-drag shift currents.
	\end{abstract}
	
	\maketitle
	\emph{Introduction---}
	The bulk photovoltaic effect (BPVE) refers to the generation of DC electric currents in uniform materials when exposed to light, without the need for heterostructures such as p-n junctions \cite{Andrewreview, Ma2023}. Unlike traditional solar cells, BPVE can generate photovoltage exceeding the material's bandgap, and its efficiency is not constrained by the Shockley–Queisser limit \cite{Shockley1961, Cook2017}. The modern theory of BPVE was developed over two decades ago \cite{Baltz1981, Sipe1995, Sipe2000}, identifying two key types of photocurrents: shift current and injection current. The shift current arises from the displacement of Wannier centers of electronic states involved in optical transitions, while the injection current results from the asymmetric injection of carriers into specific momentum states. More recently, first-principles methods have advanced the study of these photocurrents in real materials \cite{Young2012, Steve2012, Tan2016}. In the last decade, the quantum geometric aspects of BPVE have also garnered significant attention, particularly in topological materials \cite{Morimoto2016, Junyeong2020, Binghai2020, Watanabe2021, Ma2021D, Ahn2022}. However, BPVE has traditionally been associated only with noncentrosymmetric materials, leaving a large class of centrosymmetric materials largely unexplored.
	
	A promising recent approach to overcoming this limitation involves photon-drag effects \cite{JustinPRL, JustinPRB}. In these effects, the momentum of incident photons is transferred to electrons in a material. Formally, the initial state $\ket{m\bm{k}-\frac{\bm{q}}{2}}$ is excited to $\ket{n\bm{k}+\frac{\bm{q}}{2}}$, where $m, n$ denote band indices and $\bm{q}$ represents the photon momentum [see Fig.~\ref{fig:fig1}].  While this theoretical proposal is intriguing, direct experimental observations of photon-drag BPVE—such as shift and injection currents—in real materials remain elusive. Photon-drag effects have been observed in various materials, including germanium \cite{gibson1970photon}, quantum wells \cite{Luryi1987, Wieck1990}, and more recently in semimetals using THz lasers \cite{Karch2010, Jiang2011, McIver2012, Liang2023}. However, these experiments are difficult to directly associate with BPVE. One of the main challenges is that photon drag can generate multiple photocurrent contributions, making it hard to isolate specific currents. Another motivation for our work is that, unlike conventional BPVE, the relationship between photon-drag BPVE and quantum geometric tensors remains largely unexplored. Although the shift current dipole was introduced in ref.~\cite{JustinPRL}, the widely used quantum geometric quantities \cite{Resta2011}, such as the quantum metric and Berry curvature, have yet to be incorporated into this framework.
	
	In this work, we first systematically derive all possible photon-drag-induced photocurrents using the density matrix method.  Then, by explicitly expanding the formalism at small $\bm{q}$, we reveal that all of these photon-drag currents are directly related to quantum geometric tensors, as summarized in Table I. Remarkably, in centrosymmetric non-magnetic materials, we find that the polarization dependence of photon-drag-induced shift current follows the form $C\sin 2\alpha$ (where $\alpha$ is the polarization angle), while other currents follow $L_1\sin 4\alpha + L_2\cos 4\alpha + D$ [see Fig.~\ref{fig:fig1}(d)]. As a result, the shift current, which is directly linked to the quantum geometry of bands, can be fully isolated. We explicitly apply our theory to the topological insulator $1T'$-WTe$_2$.
	Finally, we propose that centrosymmetric magnetic Weyl semimetals are also an ideal platform for studying photon-drag BPVE. Our theoretical work provides a solid foundation and valuable experimental guidance for exploring photon-drag-induced BPVE in both non-magnetic and magnetic materials.
	
	\begin{figure}
		\centering
		\includegraphics[width=1\linewidth]{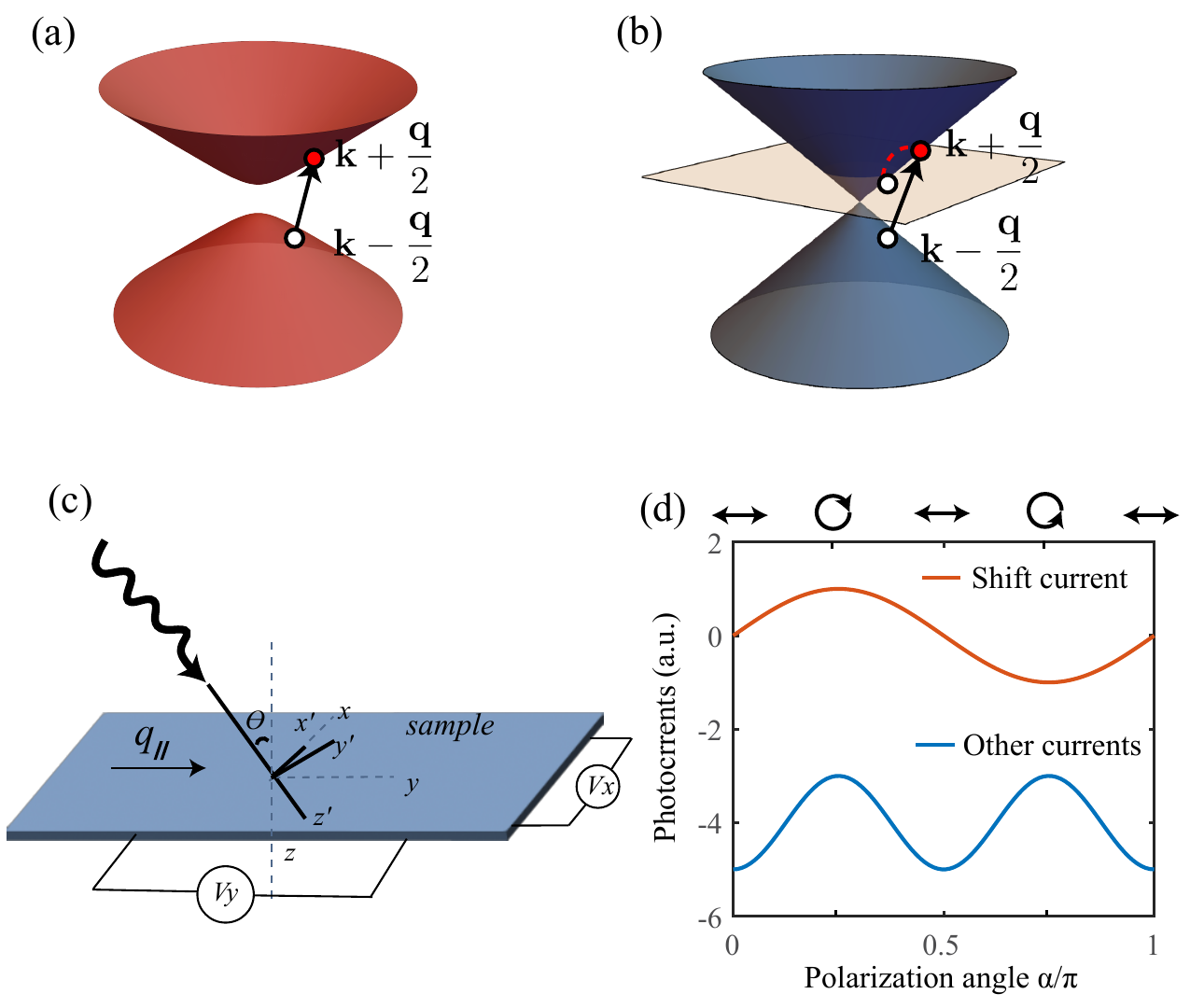}
		\caption{(a) and (b) schematically show the non-vertical photo-excitations in insulators (or semiconductors) and semimetals due to the presence of finite photon-momentum $\bm{q}$. In the presence of Fermi surfaces, the intra-band photoexcitations appear (red dashed line in (b)). (c) A light incident onto a sample with an angle $\theta$, where $xyz$ ($x'y'z'$ ) labels the sample (incident light) frame. (d) schematically show that the photon drag induced shift current follows a polarization dependence of $C\sin(2\alpha)$, while other currents show $L_1\sin(4\alpha)+L_2\cos(4\alpha)+D$ in nonmagnetic centrosymmetric materials.    }
		\label{fig:fig1}
	\end{figure}

	\emph{General formalism of photocurrents with photon-drag effects---}
	Let us first derive a general formalism for calculating photocurrents with finite photon momentum. 
	We represent the incident electromagnetic field as $\bm{E}(\bm{r},t)=\bm{E}(\omega)e^{i(\bm{q}\cdot \bm{r}-\omega t)}+\bm{E}(-\omega)e^{-i(\bm{q}\cdot \bm{r}-\omega t)}$, where $\bm{q}$ is the photon momentum and $\omega$ is the photon frequency.   We can define  the gauge potential as $A_{i}(\bm{r},t)=\sum_{\lambda}  A_{i} (\lambda \omega) e^{i\lambda(\bm{q}\cdot \bm{r}- \omega t)+\eta t}$ with $\lambda=\pm$, $\eta \rightarrow 0^{+}$ and $A_i(\lambda\omega)=\frac{ -iE_i(\lambda\omega)}{  \lambda \omega }$, so that $\bm{E}(\bm{r},t)=-\frac{\partial \bm{A}(\bm{r},t)}{\partial t}$. 

	The photocurrents can generally be derived using the density matrix method \cite{Sipe1995, Sipe2000,de_juan2020, Lingyuan2021}. The photocurrents are expressed as the trace of the density operator and the current operator, given by $\bm{j}(t)=\text{Tr}[\rho(t)\bm{\hat{J}}(t)]$.
	Here, the current operator $\bm{\hat{J}}(t)=
	\bm{\hat{J}}^{(0)}(t)+\bm{\hat{J}}^{(1)}(t)$, where $\bm{\hat{J}}^{(0)}(t)=e\frac{\partial H}{\partial \hat{\bm{p}}_i}$, and $\bm{\hat{J}}^{(1)}(t)=e^2\bm{A}_j(\bm{r},t)  \frac{\partial^2 H(\hat{\bm{p}})}{\partial \hat{\bm{p}}_{i}\partial \hat{\bm{p}}_{j}}$, with $\bm{p}$ as the momentum operator. The density matrix respects the equation of motion
	$\frac{\partial \rho(t)}{\partial t}=-\frac{i}{\hbar} [H,\rho(t)]$.
	Here,  the density matrix can be expanded in powers of electric fields $\rho(t)=\rho^{(0)}(t)+\rho^{(1)}(t)+\rho^{(2)}(t)+...$. Under the influence of an electromagnetic wave, the Hamiltonian is expressed as, $H\approx \sum_{mn,\bm{k}} \psi_{m\bm{k}}^{\dagger} [\epsilon_{n}(\bm{k})\delta_{mn}+ V_{mn}(\bm{k},t)]\psi_{n\bm{k}}$.
	Here, $m,n$ denotes the band indices, and $\epsilon_n(\bm{k})$ is the band energy.  The perturbation potential $V_{mn}(\bm{k},t)\approx V^{(1)}_{mn}(\bm{k},t)+V^{(2)}_{mn}(\bm{k},t)$. The leading order time-dependent perturbation is 
	$
	V^{(1)}_{mn}(\bm{k},t)=    e 	\sum_{\lambda} A_{i} (\lambda \omega) v^i_{m\bm{k}+\lambda \frac{\bm{q}}{2}, n\bm{k}-\lambda \frac{\bm{q}}{2}} e^{-i\lambda \omega t+\eta t}$,
	where $v^{i}_{m\bm{k}-\frac{\bm{q}}{2}, n\bm{k}+\frac{\bm{q}}{2}}=\braket{m\bm{k}-\frac{\bm{q}}{2}|\frac{\partial H}{\hbar \partial   \bm{k}_i}|n\bm{k}+\frac{\bm{q}}{2}}$ 
	represents the velocity matrix. The second order pertubation is $ V^{(2)}_{mn}(\bm{k},t)=    e^2 	\sum_{\lambda_1\lambda_2} A_{i} (\lambda_1 \omega) A_{j} (\lambda_2 \omega)v^{ij}_{m\bm{k}+\frac{(\lambda_1+\lambda_2)}{2}\bm{q},n\bm{k}-\frac{(\lambda_1+\lambda_2)}{2}\bm{q}}$
	$\times e^{-i(\lambda_1+\lambda_2) \omega t+2\eta t}$ , where 
	we define $v^{ij}_{m\bm{k}+\frac{\bm{q}}{2},n\bm{k}-\frac{\bm{q}}{2}}=\braket{m\bm{k}-\frac{\bm{q}}{2}|\frac{\partial^2 H}{\hbar^2 \partial   \bm{k}_i\partial   \bm{k}_j}|n\bm{k}+\frac{\bm{q}}{2}}$.  It can be observed that there is a finite photon momentum absorption after the photon excitations between $n$-th to $m$-th bands.  
	Finally, we can obtain 
	the second-order nonlinear DC photocurrent with
	$j^{(2)}(t)=\text{Tr}[\rho^{(1)}(t)\bm{\hat{J}}^{(1)}(t)+\rho^{(2)}(t)\bm{\hat{J}}^{(0)}(t)]$. 
	
	After careful simplifications and the application of sum rules (see Supplementary Material, SM Sec.~I) \cite{Supp}, we can decompose the photon-drag BPVE  into three distinct contributions: shift, injection (inj), and Fermi surface (fs) photocurrents (the latter arising from the presence of the Fermi surface), i.e., $j^{i}=(\sigma^{i;j,k}_{shift}+\sigma^{i;j,k}_{inj}+\sigma^{i;j,k}_{fs}) E_{j}(\omega) E_k(-\omega)
	\sigma^{i;j,k}$ with
	\begin{eqnarray}
		\sigma^{i;j,k}_{shift}&&=\frac{i\pi e^3}{\hbar^2\omega^2}\sum_{\tilde{n},\tilde{m}} f_{\tilde{n}\tilde{m}}[v^{j}_{\tilde{n},\tilde{m}}(v^{k}_{\tilde{m}, \tilde{n}})_{;k_i}-v^{k}_{\tilde{m},\tilde{n}} (v^{j}_{\tilde{n},\tilde{m}})_{;k_i}]\nonumber\\
		&&\delta(\omega_{\tilde{n}\tilde{m}}-\omega) \label{Eq_6}.\\
		\sigma^{i;j,k}_{inj}&& =\frac{-\pi e^3}{\hbar^2 \omega^2\eta } \sum_{\tilde{n},\tilde{m}} f_{\tilde{n}\tilde{m}}(v_{\tilde{n}\tilde{n}}^{i}-v_{\tilde{m}\tilde{m}}^{i})v^{j}_{\tilde{n}, \tilde{m}}v^{k}_{\tilde{m},\tilde{n}}\nonumber\\&&\delta(\omega_{\tilde{n}\tilde{m}}-\omega),\label{Eq_7}\\
		\sigma^{i;j,k}_{fs}&&=-\frac{e^3}{\hbar \omega^2} \sum_{\tilde{n},\tilde{m}} \partial_{k_i}f_{\tilde{n}\tilde{m}}  \frac{v^{j}_{\tilde{n},\tilde{m}} v^{k}_{\tilde{m}, \tilde{n}}}{\epsilon_{\tilde{n}}-\epsilon_{\tilde{m}}-\hbar\omega}.\label{Eq_8}
	\end{eqnarray}
	Here, $\omega$ is the optical frequency, $\omega_{\tilde{n}}=\epsilon_{\tilde{n}}/\hbar$, $\omega_{\tilde{n}\tilde{m}}=\omega_{\tilde{n}}-\omega_{\tilde{m}}$, $f_{\tilde{n}\tilde{m}}=f_{\tilde{n}}-f_{\tilde{m}}$, $i,j,k\in \{x,y,z\}$, and  the co-derivative of velocity matrix $(v^{j}_{\tilde{m}\tilde{n}})_{; k_i}=\partial_{k_i} v^{j}_{mn}-i (r_{\tilde{m}\tilde{m}}^{i}-r_{\tilde{n}\tilde{n}}^{i})v^{j}_{\tilde{m}\tilde{n}}$ \cite{Sipe1995} with $r_{\tilde{m}\tilde{m}}^{i}$ as intra-band Berry connection. Note that $\partial_{k_i}v_{\tilde{m}\tilde{n}}^{j}$ would include the aforementioned higher momentum contribution $v_{\tilde{m}\tilde{n}}^{ij}$ \cite{Junyeong2020}. 
	To facilitate later discussion, we define more compact notation as $\tilde{n}\equiv n\bm{k}+\frac{\bm{q}}{2}$, $\tilde{m}\equiv m\bm{k}-\frac{\bm{q}}{2}$.  It is worth noting that intra-band contributions ($m= n$) may arise when $\bm{q}\neq 0$  for the metallic case [see Fig.~\ref{fig:fig1}(b)].  For the case where $m\neq n$,   the shift and injection currents can be expressed in terms of the interband component of the position operator: $
	r_{nm}(\bm{k})=\frac{ \hbar v_{nm}(\bm{k})}{i(\epsilon_{n}(\bm{k})-\epsilon_{m}(\bm{k}))}$, which reproduces results from previous studies at $\bm{q}=0$ \cite{Sipe2000, Lingyuan2021}. The third conductivity tensor $\sigma_{fs}^{i;j,k}$ is a Fermi surface contribution but includes inter-band excitations.   While the above formulation is intuitively expected, our derivations provide a robust theoretical foundation for all photon-drag photocurrent contributions involving an arbitrary number of bands. Furthermore, it is known that shift currents can be derived using the Floquet formalism \cite{Morimoto2016}. As explicitly demonstrated in SM Sec. II, the shift current formalism incorporating photon drag [Eq.~\eqref{Eq_6}] is consistent with results obtained through the Floquet method.

	\emph{Quantum geometric nature of photon-drag BPVE}--- To extract the quantum geometric nature of BPVE via photon drag, we can expand Eqs.~\eqref{Eq_6} to \eqref{Eq_8} up to leading order in $\bm{q}$ as $\sigma^{i;j,k}(\bm{q})\approx \sigma^{i;j,k}(\bm{q}=0)+q_{\lambda}\sigma^{\lambda ijk}$. As shown in SM Sec.V \cite{Supp}, we find
	\begin{eqnarray}
		\sigma^{\lambda ijk}_{shift}&&\approx \frac{\pi e^3 }{2 \hbar^2\omega} \sum_{m\neq n} f_{nm}W_{nm}^{k} [(R^{i;\lambda}_{mn}-R^{i;j}_{nm})G^{\lambda j}_{mn}\nonumber\\
		&&+i\partial_{k_i} G_{mn}^{\lambda j}]\delta(\omega_{nm}-\omega)+(j \leftrightarrow k)^*, \label{Eq:shift_expand}
	\end{eqnarray}
	\begin{eqnarray}
		\sigma^{\lambda ijk}_{inj}&& \approx -\frac{\pi e^3 }{\hbar^2 \eta } \sum_{m\neq n} f_{nm} [\frac{1}{2\omega}\Delta_{nm}^{i} (W_{nm}^{j}G_{mn}^{k\lambda}\nonumber\\
		&&+W_{nm}^{k}G_{mn}^{\lambda j})-\frac{1}{2}\Delta_{nm}^{i} \partial_{k_\lambda} G_{mn}^{kj}]\delta(\omega_{nm}-\omega),
	\end{eqnarray}
	\begin{eqnarray}
		\sigma^{\lambda ijk}_{fs}&&\approx \frac{e^3 }{2\hbar^2 \omega^2} \sum_{m\neq n} f_{nm}[\partial_{k_i} (G_{mn}^{k\lambda})\frac{W_{nm}^{j}\epsilon_{nm}}{\epsilon_{nm}-\hbar\omega}\nonumber\\
		&&+G_{mn}^{k\lambda}\partial_{k_i}(\frac{W_{nm}^{j}\epsilon_{nm}}{\epsilon_{nm}-\hbar\omega}) +(j\leftrightarrow k)^{*}].
	\end{eqnarray}
	Here, $W_{nm}^{k}=v_{nn}^{k}+v_{mm}^{k}$, $\Delta_{nm}^{i}=v_{nn}^{i}-v_{mm}^{i}$, $R_{nm}^{i;j}=r^i_{nn}-r_{mm}^{i}+i\partial_{k_i}\log(r_{nm}^j)$ is the shift vector, the quantum geometry tensor: $G_{mn}^{jk}=r_{mn}^{j}r_{nm}^{k}=\mathcal{G}_{mn}^{jk}-\frac{i}{2}\Omega_{mn}^{jk}$ with the quantum metric tensor $\mathcal{G}_{mn}^{jk}=\text{Re}[r^j_{mn}r^k_{nm}]$, and Berry curvature tensor $\Omega_{mn}^{jk}=-2\text{Im}[r^j_{mn}r^k_{nm}]$. Note that only inter-band contributions ($m\neq n$) are considered here, and we have neglected the three-band process [see SM Sec.IV for detailed approximations]. Also, the above expansion is valid only for insulators or semiconductors where the conduction and valence bands do not intersect.   When $j=k$, all $\sigma^{\lambda ijk}$ tensors are real, corresponding to linearly polarized light-driven photocurrents. For $j\neq k$, the imaginary part of $\sigma^{\lambda ijk}$ is non-zero,  corresponding to circularly polarized light-driven photocurrents. Remarkably, as summarized in Table I,   the photon drag induced linear (L)-shift currents are directly related to the quantum metric $\mathcal{G}$ and Berry curvature dipole $\partial_{k}\Omega$, while photon drag induced circular(C)-shift currents are directly related to the Berry curvature $\Omega$ and quantum metric dipole $\partial_{k}\mathcal{G}$.  Note that we directly capture the geometry nature of photon-drag shift currents with quantum geometry tensor rather than introducing a shift current dipole as ref.~\cite{JustinPRL}.   On the other hand, the quantum metric $\mathcal{G}$ and its dipole $\partial_{k}\mathcal{G}$  are associated with L-injection/fermi surface part of photon-drag currents, while the C-injection/fermi surface contribution is related to Berry curvature $\Omega$ and its dipole $\partial_{k}\Omega$.  The carrier dynamics play a crucial role in both the injection and Fermi surface currents, whereas the change in the Wannier center during optical excitations is essential for the shift current. Consequently, we observe that the geometric characteristics of the Fermi surface photocurrent align with the behavior of an injection current rather than a shift current.
	Furthermore, to apply the above analysis to nonmagnetic centrosymmetric systems,  the Hamiltonian can be divided into two blocks related by time-reversal symmetry, ensuring that the quantum geometry tensor remains well-defined within each block.   
	
	\begin{table}
		\caption{Quantum geometric nature of photon-drag  BPVE.}
		\centering
		\begin{tabular}{c|c|c|c|c|}
			\hline\hline
			Photon-drag response    &   L-shift  & C-shift& L-inj/fs& C-inj/fs\\\hline
			Geometry quantity     & $\mathcal{G}, \partial_{k}\Omega$   &$\Omega, \partial_{k}\mathcal{G}$ &$\mathcal{G}$, $\partial_{k}\mathcal{G}$& $\Omega$, $\partial_{k}\Omega$ \\\hline
			T condition    & \xmark &--- & ---& \xmark \\ \hline
		\end{tabular}
		\label{TableI}
	\end{table}

	
	Another interesting observation is that the particle-hole asymmetry, characterized by $W_{nm}^{k}=v_{nn}^{k}+v_{mm}^{k}$, plays a crucial role in enhancing photon-drag effects. According to the form of Eq.~\eqref{Eq:shift_expand},  the photon-drag shift current is fully suppressed when  $W_{nm}^{k}=0$. Physically, in the leading order perturbation,   there are two paths going from $m\bm{k}-\frac{\bm{q}}{2}$ to  $n\bm{k}+\frac{\bm{q}}{2}$: (i) a vertical interband transition followed by intraband transition, or (ii) an intraband transition followed by a vertical interband transition [see paths 1 and 2 in Fig.~\ref{fig:fig2}(a)].  The photocurrents generated by these two paths lead to the factor $W_{nm}^{k}=v_{nn}^{k}+v_{mm}^{k}$, which is zero when $v_{nn}^{k}$ and  $v_{mm}^{k}$ are opposite in sign.

	\emph{Gauge invariance and symmetry properties of photon-drag BPVE---} After highlighting the geometry nature of the photon-drag photocurrent formalism Eq.~\eqref{Eq_6} to \eqref{Eq_8}, we now examine their gauge invariance and symmetry properties.  Gauge invariance is not immediately apparent when both time-reversal and inversion symmetry are present, as the bands are doubly degenerate. In this situation, we can relabel $\tilde{n}\mapsto \tilde{n}\alpha_n$ and $\tilde{m}\mapsto \tilde{m}\alpha_m$, where $\alpha_{n,m}\in \{1,2\}$ labels the doubly degenerate bands. We also need to generalize the co-derivative as 
	$
	(v^{j}_{\tilde{n}\alpha_n,\tilde{m}\alpha_m})_{;k_i}= \partial_{k_i} v^{j}_{\tilde{n}\alpha_n,\tilde{m}\alpha_m}-i\sum_{\alpha} (r^{i}_{\tilde{n}\alpha_n,\tilde{n}\alpha}v^{j}_{\tilde{n}\alpha, \tilde{m}\alpha_m}-v^{j}_{\tilde{n}\alpha_n,\tilde{m}\alpha}r^{i}_{\tilde{m}\alpha, \tilde{m}\alpha_{m}}).$
	Here, $r^{i}_{\tilde{n}\alpha_{n},\tilde{m}\alpha_{m}}=\braket{\tilde{n}\alpha_{n}|i\nabla_{\bm{k}}|\tilde{m}\alpha_{m}}$ is the Berry connection.
	Under a $U(2)$ gauge transformation,
	$v^{j}\mapsto U^{\dagger}v^{j}U$ and $r^{i}\mapsto U^{\dagger}r^{i}U+i U^{\dagger} \partial_{k_i} U$. Using $U^{\dagger}U=1$, it is straightforward to show that $\sum_{\alpha_m,\alpha_n} v^{k}_{m\alpha_m,n\alpha_n} (v^{j}_{n\alpha_n,m\alpha_m})_{;k_i}$ and $\sum_{\alpha_m,\alpha_n} v^{j}_{n\alpha_n,m\alpha_m} v^{k}_{m\alpha_m,n\alpha_n}$  are invariant under this gauge transformation. For further details, see SM III.  Note that the energy and Fermi distributions are gauge invariant. As a result, we conclude that the nonlinear photoconductivity  Eq.~\eqref{Eq_6}-\eqref{Eq_8} remains gauge invariant in the presence of a doubly degenerate band.

	Next, we discuss the symmetry properties of the conductivity tensor $\sigma^{i;j,k}(\bm{q})$. The most relevant symmetries for our analysis are time-reversal and inversion symmetry. In the case of zero photon momentum ($\bm{q}=0$), the entire conductivity tensor changes sign under inversion.   Under time-reversal symmetry, the real part of the shift current tensor flips sign, while the imaginary part does not. Conversely, for the injection/Fermi surface current tensor, the real part remains unchanged, while the imaginary part flips the sign. The presence of finite photon momentum $\bm{q}$ would change to -$\bm{q}$ under time-reversal or inversion operation. As a result, we expect that under inversion symmetry: 
	\begin{equation}
		\sigma_{shift/inj/fs}^{i;j,k}(\bm{q})=-\sigma_{shift/inj/fs}^{i;j,k}(-\bm{q}).
	\end{equation}
	In this case, all second-order conductivity tensors vanish when $\bm{q}=0$. In the presence of time-reversal symmetry, the following relations hold:
	\begin{eqnarray}
		\sigma_{shift,p}^{i;j,k}(\bm{q})&&=(-1)^{p}\sigma_{shift,p}^{i;j,k}(-\bm{q}), \label{Eq_13}\nonumber\\
		\sigma_{inj/fs,p}^{i;j,k}(\bm{q})&&=-(-1)^{p}\sigma_{inj/fs,p}^{i;j,k}(-\bm{q})\label{Eq_14}\label{Eq_15},
	\end{eqnarray}
	where $(-1)^{p}$ is 1 for the real part ($p=R$) and -1 for the imaginary part ($p=I$), corresponding to linearly polarized and circularly polarized light respectively.  These symmetry relations indicate that with time-reversal symmetry, the shift current induced by photon-drag must involve circularly polarized light, whereas the injection current can arise with linearly polarized light. Moreover, the injection and Fermi surface photocurrents follow the same symmetry rules. The mixing between injection and Fermi surface photocurrents occurs only in metallic systems, as Fermi surface photocurrents vanish in insulating systems. A summary of the symmetry-allowed photon-drag-induced currents under inversion (P) and time-reversal (T) symmetry is provided in Table I.

	\begin{figure}
		\centering
		\includegraphics[width=1\linewidth]{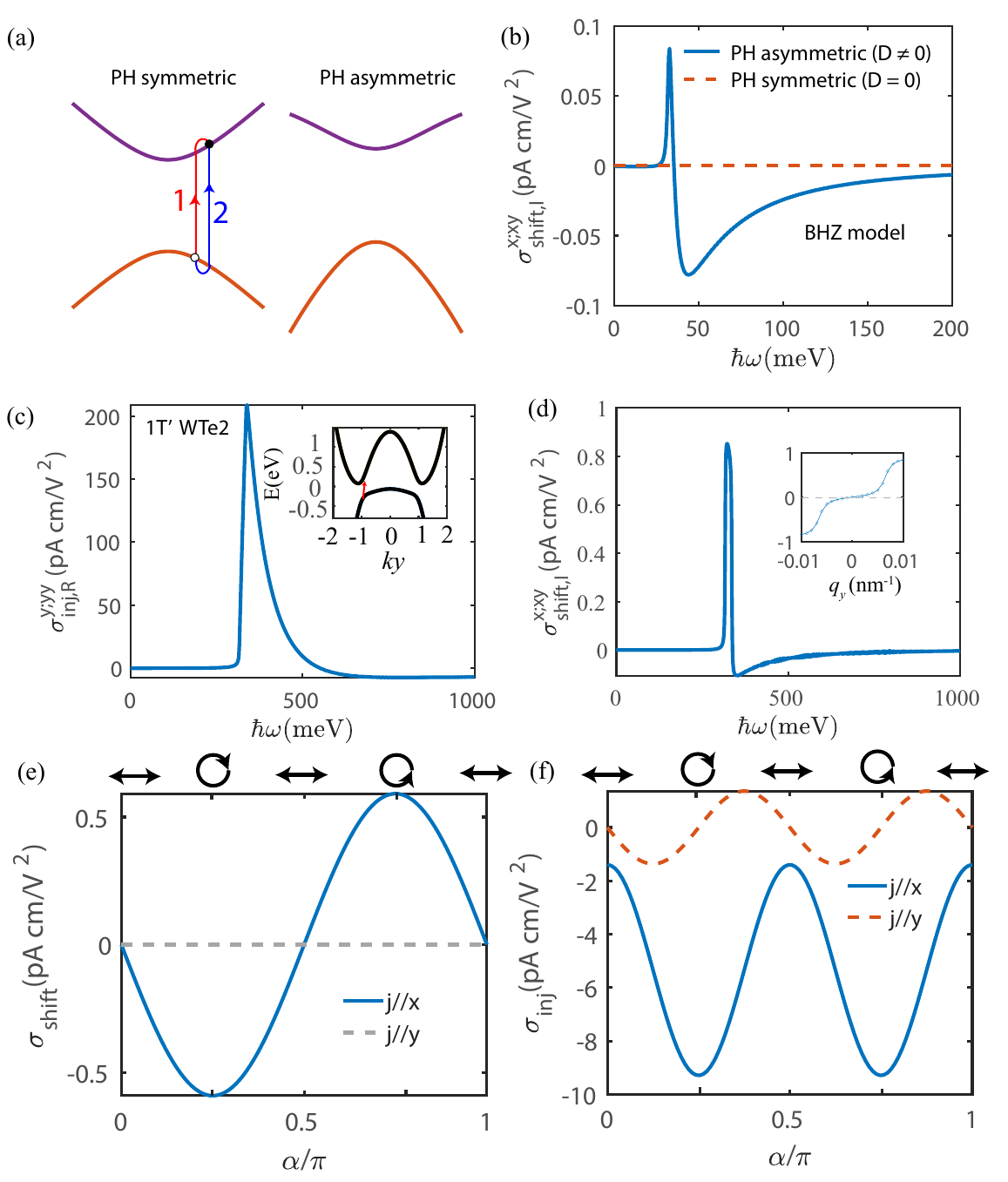}
		\caption{ (a) illustrates the bands with and without particle-hole (PH) symmetry. The blue line and red line highlight two distinct excitation paths. (b) The photon-drag shift conductivity tensor  $\sigma_{shift, I}^{x;xy}(\bm{q}_{\parallel})$ from BHZ model with PH ($D=0$) and without PH symmetry ($D=-0.594$ eV). Other parameters are $A=-3.62$ eV, $B=-18$ eV, $C=-0.018$ eV, $M=0.00922$ eV \cite{Bernevig2013}.  (c), (d) The calculated photon-drag shift and injection conductivity tensor $\sigma_{inj, R}^{y;yy}$  and $\sigma_{shift, I}^{x;xy}$ versus 
			photon energy in 1T$'$ WTe$_2$. The inset of (c) shows the band structure of 1T$'$ along (0,$k_y$), while the inset of (d) is to highlight the $q$-dependence of the conductivity tensor with $\omega=320$ meV. (e), (f) display the total shift and injection current polarization dependence of the $x$ and $y$ direction, respectively.   Without specific mention in this work, we fix the photon momentum along $y$-direction with $q=0.01$ nm$^{-1}$, the incident angle $\theta=45^{\circ}$ in the numerical calculations.   }
		\label{fig:fig2}
	\end{figure}
	
	\emph{Polarization dependence of photocurrents---}
	In the previous section, we identified the types of photon-drag photocurrents in centrosymmetric crystals, depending on the polarization of light.  Actually,  the linearly polarized light can be continuously tuned into circularly polarized light by rotating the angle $\alpha$ of a $\lambda/4$ waveplate. In this case, the electric field of light is written as $\bm{E}(t)=|E|e^{i\bm{q}\cdot \bm{r}-i\omega t}[(\cos^2\alpha+i\sin^2\alpha)\hat{x'}+(\sin\alpha\cos\alpha (1-i)) \hat{y'}]+c.c.$ \cite{Liang2023} with $\bm{q}=q\hat{z'}$ and $\alpha$ as the light polarization angle, where the light polarization varies at a $180^{\circ}$ (LP at $\alpha=0^\circ$, right-handed CP at $\alpha=45^\circ$, LP at $\alpha=90^\circ$, left-handed CP at $\alpha=135^\circ$, LP at $\alpha=180^\circ$).  As shown in the Fig.~\ref{fig:fig1}(c), the basis vectors of the light frame $\hat{x'}\hat{y'}\hat{z'}$ 
	and the sample frame  $\hat{x}\hat{y}\hat{z}$  are related through $\hat{x'}=\hat{x},
	\hat{y'}=\cos\theta \hat{y}-\sin\theta\hat{z},
	\hat{z'}=\sin\theta \hat{y}+\cos\theta \hat{z}$. The photon momentum direction in the sample frame is 
	$\bm{q}=q\hat{z'}=q\sin\theta\hat{y}+q\cos\theta \hat{z}$.
	We now assume that the photocurrents are confined to the $xy$-plane, meaning that only $E_x$, $E_y$ and  $q_{\parallel}=q\sin\theta$ are relevant [see an illustration of the experimental setup in Fig.~\ref{fig:fig1}]. Note the incident angle $\theta$ must be non-zero in order to induce photon-drag within the $xy$ plane.   Using the relation between the light frame and sample frame, we find that the polarization dependence of  photocurrents is given by 
	\begin{eqnarray}
		&& j^{i}=[(1-\frac{1}{2}\sin^22\alpha)\sigma_{R}^{i;xx}(\bm{q}_{\parallel})+ \frac{\cos^2\theta\sin^22\alpha}{2}\sigma_{R}^{i;yy}(\bm{q}_{\parallel})\nonumber\\
		&&+\frac{\cos\theta}{2}\sin(4\alpha)\sigma_{R}^{i;xy}(\bm{q}_{\parallel})-\cos\theta \sin(2\alpha)\sigma_{I}^{i;xy}(\bm{q}_{\parallel})]|E|^2\nonumber
	\end{eqnarray}
	This simplifies to:
	\begin{equation}
		j^{i}= C\sin 2\alpha+L_1\sin 4\alpha+L_2\cos 4\alpha+ D. \label{Eq_polar}
	\end{equation}
	where $C$ characterizes helicity-dependent photocurrents,  $L_1$ and $L_2$ are coefficients capturing linear polarization dependence of the photocurrents, and $D$ is a polarization-independent contribution.
	
	In a centrosymmetric nonmagnetic crystal with photon-drag effects, as we analyzed, the imaginary part of the conductivity tensor arises from shift currents, while the real part is due to the injection and Fermi surface photocurrents. Based on the form of Eq.~\eqref{Eq_polar}, we find that the polarization dependence of shift currents is $C\sin(2\alpha)$, while the one of injection and Fermi surface photocurrents is  $L_1\sin 4\alpha+L_2\cos 4\alpha+ D$. This means that shift currents can be completely separated from injection and Fermi surface photocurrents by analyzing their distinct polarization dependencies. 
	\begin{figure}[b]
		\centering
		\includegraphics[width=1\linewidth]{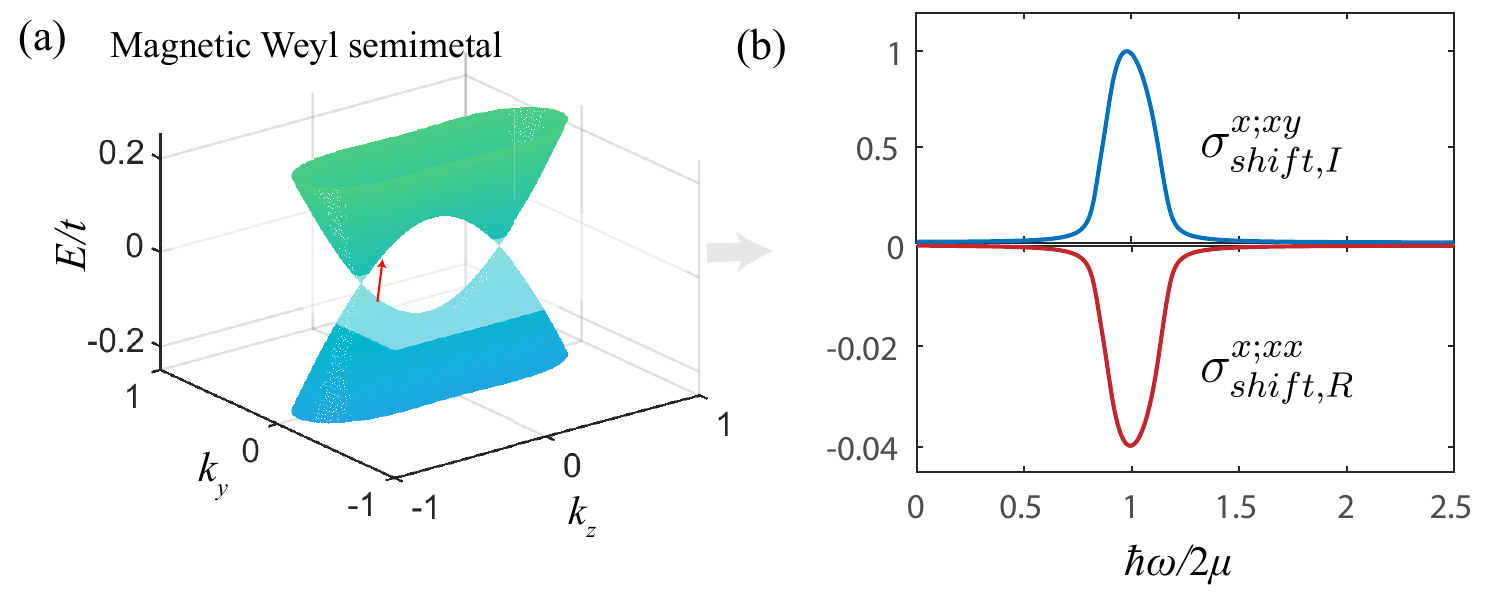}
		\caption{The photon drag induced photovoltaic effects in a 3D magnetic Weyl semimetal. (a) shows the band structures of the magnetic Weyl metal.. (b) The interband circular, linear shift conductivity tensor $\sigma_{shift, I}^{x;xy}$,  $\sigma_{shift, R}^{x;xx}$ versus photon energy. Both curves have normalized the maximum of $\sigma_{shift, I}^{x;xy}$. In this figure, we adopt the parameters $t=t_x=t_y=t_z=1$ eV, $\gamma=0.9$. }
		\label{fig:fig3}
	\end{figure}
	
	\emph{Applicaiton to nonmagnetic topological insulators---} We will first numerically demonstrate the aforementioned particle-hole symmetry argument using the Bernevig-Hughes-Zhang (BHZ) model: $H_{BHZ}(\bm{k})=\epsilon_0(\bm{k})s_0\tau_0+M(\bm{k})s_0\tau_z+Ak_xs_z\tau_x+Ak_ys_0\tau_y$, where $s, \tau$ are Pauli matrix operating in spin and orbital space, respectively.
	Here, $\epsilon_0(\bm{k})=C-D(k_x^2+k_y^2)$,  $M(\bm{k})=M-B(k_x^2+k_y^2)$.  The parameter 
	$D$ controls the particle-hole asymmetry. In the following numerical calculation,  we would fix the light frame and sample frame as shown in Fig.~\ref{fig:fig1}(c), such that the photon momentum $\bm{q}_{\parallel}$ is along $y$ direction. Our calculations focus on the bulk optical transitions. The possible photocurrents arising from the excitations between the edge states and bulk states are not considered there \cite{Dantscher2017}, assuming the light spots are away from the sample edges.  Numerically, we calculate the photon-drag circular shift conductivity tensor $\sigma^{x;xy}_{shift,I}$ using Eq.~\eqref{Eq_6}. As shown in Fig.~\ref{fig:fig2}(b), the photon-drag shift currents are suppressed when $D$ is set to zero, i.e.,  when particle-hole symmetry is restored, which aligns with our theoretical expectations.

	There is a realistic topological insulator material that exhibits prominent particle-hole asymmetry:  the topological monolayer 1T$'$ WTe$_2$ \cite{Xiaofeng,Sanfeng, Tang2017,Fei2017}. To generate photocurrents in this centrosymmetric system, previous experiments applied an out-of-plane displacement field to break inversion symmetry \cite{Xu2018, Ma2019}. In contrast, we now employ photon drag to enable photocurrents in this nonmagnetic centrosymmetric system.  Specifically, we take the low-energy effective Hamiltonian of the monolayer 1T$'$ WTe$_2$ 
	from ref.~\cite{Law2020}, which leads to the band structures shown in the inset of Fig.~\ref{fig:fig2}(c).   The system respects time-reversal symmetry $T$,  inversion symmetry $P$, and a mirror symmetry $M_y$ that maps $y$ to $-y$.  Due to these symmetries, it is straightforward to deduce that only four photon-drag conductivity tensors are non-zero in this centrosymmetric material when $\bm{q}_{\parallel}\neq 0$: $\sigma^{x;xy}_{shift, I} (\bm{q}_{\parallel})$,  $\sigma^{y;xx}_{inj,R}(\bm{q}_{\parallel})$, $\sigma^{x;xy}_{inj,R}(\bm{q}_{\parallel})$, and $\sigma^{y;yy}_{inj,R}(\bm{q}_{\parallel})$. 
	As an illustration, the calculated frequency dependence of $\sigma_{inj,R}^{y;yy}$ and $\sigma_{shift,I}^{x;xy}$  is plotted Figs.~\ref{fig:fig2} (c) and (d).   Note that all these photocurrent tensors in this centrosymmetric sample are finite only when  $\bm{q}_{\parallel}\neq 0$ [illustrated in the inset of Fig.~\ref{fig:fig2}(d)].  
	
	The photon-drag shift current proposed in WTe$_2$ is also experimentally observable. The optimal magnitude of photon-drag shift tensor can reach the order of 0.7 pAcm/V$^2$ in WTe$_2$, which corresponds to a 2D photoresponsivity of $\kappa_{2D}\sim 0.5$ nA cm/W \cite{Junyeong2020}. The photoresponsivity in this scale can be easily observed through standard photocurrent measurements \cite{Zhang2019I}.   Moreover,  in the SM Sec.VII,  the zero-q shift conductivity tensor is obtained by breaking the inversion symmetry in the WTe$_2$ model. It is observed that the peak values of the photon-drag shift photocurrent and the noncentrosymmetric shift photocurrent can be of the same order. 

	
	The polarization dependence of the photon-drag photocurrents in 1T$'$ WTe$_2$ can be obtained using  Eq.~\eqref{Eq_polar}. As shown in Fig.~\ref{fig:fig2} (e) and (f), the shift and injection currents exhibit the expected polarization dependence of  $C\sin(2\alpha)$ and $L_1\sin4\alpha+ L_2\sin4\alpha+D$, respectively. It is also worth noting that the shift current is finite only along $x$-direction. These observations can be experimentally verified by measuring the polarization dependence of the photocurrents and decoupling them into different polarization-dependent channels \cite{McIver2012}.


	\emph{photon-drag BPVE in centrosymmetric magnetic Weyl semimetals---} The nonlinear optical responses have been extensively explored in noncentrosymmetric Weyl semimetals \cite{Wu2017,deJuan2017,Patrick2017,Sundong2019,Moore2021, Ahn2022,Cook2024}.  Next, we demonstrate that the centrosymmetric weyl semimetals also serve as excellent platforms for exploring photon-drag BPVE.  Specifically, we use the magnetic Weyl model from ref.~\cite{Weyl2018}:
	\begin{eqnarray}
		H_{\text{Weyl}}(\bm{k})&&=t_z(2-\cos k_x-\cos k_y-\cos k_z+\gamma)\nonumber\\
		&&+t_x \sin k_x \tau_x+ t_y\sin k_y \tau_y-\mu.
	\end{eqnarray}
	where the Pauli matrices $\tau$ are defined in orbitals with opposite parity (such as $s,p$ orbitals), and the weyl points locates at $(0,0, \pm k_0)$ with $\cos k_0=\gamma$ [see Fig.~\ref{fig:fig3}(a)]. It breaks time-reversal $T=K$ ($K$ is complex conjugate), but preserves inversion symmetry $P=\tau_z$ with $PH(\bm{k})P^{-1}=H(-\bm{k})$.

	The calculated circular shift conductivity tensor $\sigma_{shift, I}^{x;xy}$ and linear shift conductivity tensor $\sigma_{shift, R}^{x;xx}$ under photon drag are shown in Fig.~\ref{fig:fig3}(b). Both are finite and exhibit a peak near $\hbar\omega/2\mu$ due to the presence of Weyl points, where $\mu$ is the chemical potential measured from Weyl points.  Interestingly, in this case, shift currents can be excited in centrosymmetric materials via photon drag, even with linearly polarized light, due to the breaking of time-reversal symmetry.  Note that the earlier particle-hole symmetry argument does not apply to Weyl semimetals, as the two bands touch at the Weyl points, violating the assumptions used to derive Eq.~\ref{Eq:shift_expand}.


	\emph{Conclusion and Discussion---} In conclusion, we have presented a comprehensive theoretical analysis of photon-drag BPVE. In particular, we have identified the quantum geometric nature of photon-drag BPVE.  Additionally, we emphasized that the photon-drag shift currents can be distinguished through their polarization dependence in nonmagnetic centrosymmetric materials. These findings will offer valuable theoretical and experimental insights for future studies on photon-drag BPVE.

	We summarize key aspects for enhancing photon-drag shift responses based on our theory. In addition to symmetry requirements:
	(i) The magnitude and incident angle of the photon momentum play a crucial role. It has been suggested that photon momentum can be amplified by coupling with polaritons \cite{JustinPRL, Kurman2018};
	(ii) Since the photon-drag shift current effect is closely tied to the quantum geometric tensor, multiband materials with significant quantum geometric contributions from their energy bands—such as topological insulators or semimetals—are expected to enhance this effect;
	(iii) For insulating materials, as demonstrated in Fig. 2, the asymmetry between conduction and valence bands further enhances photon-drag shift currents.
	
	In the main text, we primarily applied our theory to $Z_2$ 
	topological insulators. Our theory can be readily applied to topological crystalline materials, such as SnTe material class\cite{Hsieh2012}, where the quantum geometric properties of the band structure are equally prominent.
	Moreover, our theory of photon-drag BPVE can also be applied to noncentrosymmetric materials. Recently, however, photon drag-induced bulk photovoltaic effects have also garnered attention in noncentrosymmetric materials \cite{Ji2024}. We would like to emphasize that our theoretical framework, including the quantum geometric analysis, is also applicable in these cases.

	Another intriguing platform to explore photon-drag shift current effects is topological materials with nontrivial magnetoelectric polarizability, such as finite-size topological systems \cite{Crook2023finite} or axion insulators \cite{Moore2009}. The underlying insight is that the photon-drag photocurrent arises from the generation of current under the dynamic electric and magnetic fields of light. In materials with magnetoelectric polarizability, the magnetic component of light is expected to influence the electric polarization, while the time-dependent variation of polarization would directly generate a photocurrent. Consequently, the systems with nontrivial magnetoelectric polarizability may exhibit nontrivial photon-drag contributions.

	\emph{Acknowledgments---} Y.M.X.  acknowledges financial support from the RIKEN Special Postdoctoral Researcher(SPDR) Program.
	N.N. was supported by JSPS KAKENHI Grant No. 24H00197 and 24H02231.
	N.N. was also supported by the RIKEN TRIP initiative.  
	
	\emph{Data Availability---} All study data are included in the article and/or SI Appendix.
	
	\emph{Competing interests---} The authors declare no competing interest.
	

%

 \clearpage
	
	\onecolumngrid
	\begin{center}
		\textbf{\large Supplementary Material for  \lq\lq{} Photon-drag photovoltaic effects and quantum geometric nature \rq\rq{}}\\[.2cm]
		Ying-Ming Xie,$^{1}$ Nato Nagaosa,$^{1,2}$ \\[.1cm]
		{\itshape ${}^1$    RIKEN Center for Emergent Matter Science (CEMS), Wako, Saitama 351-0198, Japan}\\
			{\itshape ${}^2$   Fundamental Quantum Science Program, TRIP Headquarters, RIKEN, Wako 351-0198, Japan}\\[1cm]
	\end{center}
	\setcounter{equation}{0}
	\setcounter{section}{0}
	\setcounter{figure}{0}
	\setcounter{table}{0}
	\setcounter{page}{1}
	\renewcommand{\theequation}{S\arabic{equation}}
	\renewcommand{\thetable}{S\arabic{table}}
	\renewcommand{\thesection}{\Roman{section}}
	\renewcommand{\thefigure}{S\arabic{figure}}
	\renewcommand{\bibnumfmt}[1]{[S#1]}
	\renewcommand{\citenumfont}[1]{#1}
	\makeatletter
	
	\onecolumngrid
	
	\maketitle
\section{Bulk photovoltaic effects with photon drag}
\subsection{Photocurrents from density matrix formalism method}
Let us consider the incident electromagnetic field as $\bm{E}(\bm{r},t)=\bm{E}(\omega)e^{i(\bm{q}\cdot \bm{r}-\omega t)}+\bm{E}(-\omega)e^{-i(\bm{q}\cdot \bm{r}-\omega t)}$.   We can define $A_{i}(\bm{r},t)=\sum_{\lambda}  A_{i} (\lambda \omega) e^{i\lambda(\bm{q}\cdot \bm{r}- \omega t)+\eta t}$ with $A_i(\lambda\omega)=\frac{ -iE_i(\lambda\omega)}{  \lambda \omega }$ so that $\bm{E}(\bm{r},t)=-\frac{\partial \bm{A}(\bm{r},t)}{\partial t}$. Here $\eta \rightarrow 0^{+}$ so that the perturbation is turned on adiabatically from $t\rightarrow -\infty$. It is important to note that the presence of finite $\bm{q}$ indicates that we take both the electric and magnetic fields of light. The magnetic field component of light is given by
$\bm{B}(\bm{r},t)
=\frac{\bm{q}\times \bm{E}(\bm{r},t)}{\omega}$. The presence of both $\bm{B}$ and $\bm{E}$ fields so that the photon drag effects was initially recognized as a Hall effect.
In the vector gauge, the Hamiltonian under the electromagnetic wave is given by
\begin{eqnarray}
	H(\hat{\bm{p}}+e\bm{A})&&\approx  	H(\hat{\bm{p}})+V^{(1)}(\bm{r},t)+V^{(2)}(\bm{r},t),\\
	V^{(1)}(\bm{r},t)&&=	e\bm{A}(\bm{r},\bm{t})\cdot \frac{\partial H(\hat{\bm{p}})}{\partial \hat{\bm{p}}}\\
	V^{(2)}(\bm{r},t)=	&&e^2\bm{A}_{i}(\bm{r},t)\bm{A}_{j}(\bm{r},t)\cdot \frac{\partial^2 H(\hat{\bm{p}})}{\partial \hat{\bm{p}}_{i}\partial \hat{\bm{p}}_{j}}
\end{eqnarray}
The current operator up to the second order is
\begin{equation}
	\hat{\bm{J}}_i=\frac{\delta H}{\delta \bm{A}}=\bm{\hat{J}}^{(0)}(t)+\bm{\hat{J}}^{(1)}(t).
\end{equation}
Here, $\bm{\hat{J}}^{(0)}(t)=e\frac{\partial H}{\partial \hat{\bm{p}}_i}$, and $\bm{\hat{J}}^{(1)}(t)=e^2\bm{A}_j(\bm{r},t)  \frac{\partial^2 H(\hat{\bm{p}})}{\partial \hat{\bm{p}}_{i}\partial \hat{\bm{p}}_{j}}$.
In the band basis
\begin{equation}
	H=\sum_{mn} \psi_{m\bm{k}}^{\dagger} [\epsilon_{n}(\bm{k})\delta_{mn}+ V^{(1)}_{mn}(\bm{k},t)+V^{(2)}_{mn}(\bm{k},t)]\psi_{n\bm{k}}.
\end{equation}
where the time-dependent perturbation is
\begin{eqnarray}
	V^{(1)}_{mn}(\bm{k},t)&&=    e 	\sum_{\lambda} A_{i} (\lambda \omega) v^i_{m\bm{k}+\lambda \frac{\bm{q}}{2}, n\bm{k}-\lambda \frac{\bm{q}}{2}} e^{-i\lambda \omega t+\eta t},\\
	V^{(2)}_{mn}(\bm{k},t)&&=    e^2 	\sum_{\lambda_1\lambda_2} A_{i} (\lambda_1 \omega) A_{j} (\lambda_2 \omega)v^{ij}_{m\bm{k}+\frac{(\lambda_1+\lambda_2)}{2}\bm{q},n\bm{k}-\frac{(\lambda_1+\lambda_2)}{2}\bm{q}} e^{-i(\lambda_1+\lambda_2) \omega t+2\eta t}.
\end{eqnarray} 
The velocity matrices 
\begin{eqnarray}
	v^{i}_{m\bm{k}+\frac{\bm{q}}{2}, n\bm{k}-\frac{\bm{q}}{2}}=\braket{m\bm{k}+\frac{\bm{q}}{2}|\frac{\partial H}{\hbar \partial   \bm{k}_i}|n\bm{k}-\frac{\bm{q}}{2}},\\
	v^{ij}_{m\bm{k}+\frac{\bm{q}}{2},n\bm{k}-\frac{\bm{q}}{2}}=\braket{m\bm{k}+\frac{\bm{q}}{2}|\frac{\partial^2 H}{\hbar^2 \partial   \bm{k}_i\partial   \bm{k}_j}|n\bm{k}-\frac{\bm{q}}{2}}.
\end{eqnarray}
The equation of motion for the density matrix is
\begin{equation}
	\frac{\partial \rho(t)}{\partial t}=-\frac{i}{\hbar} [H,\rho(t)]
\end{equation}
We can expand the density matrix in powers of $\bm{E}$:
\begin{equation}
	\rho(t)=\rho^{(0)}(t)+\rho^{(1)}(t)+\rho^{(2)}(t)+...
\end{equation}
Here, $\rho^{(0)}(t)$ is the zeroth order density matrix in the absence of external perturbation, i.e., the Fermi distribution function with $\rho_{mn}^{(0)}(t)=\delta_{mn} f_m(\bm{k})$.
The first-order density matrix is given by
\begin{equation}
	\frac{\partial \rho^{(1)}(t)}{\partial t}=-\frac{i}{\hbar} [H^0,\rho^{(1)}(t)]-\frac{i}{\hbar} [V^{(1)},\rho^{(0)}]
\end{equation}
where $H^0$ is the unperturbed Hamiltonian. Similarly,  the second-order density matrix is given by
\begin{equation}
	\frac{\partial \rho^{(2)}(t)}{\partial t}=-\frac{i}{\hbar} [H^0,\rho^{(2)}(t)]-\frac{i}{\hbar} [V^{(1)},\rho^{(1)}]-\frac{i}{\hbar} [V^{(2)},\rho^{(0)}].
\end{equation}
The trace of the density operator and the current operator would give the photocurrents:
\begin{equation}
	\bm{j}(t)=\text{Tr}[\rho(t)\bm{\hat{J}}(t)].
\end{equation}
The second-order nonlinear contribution is
\begin{equation}
	j^{(2)}(t)=	\text{Tr}[\rho^{(1)}(t)\bm{\hat{J}}^{(1)}(t)+\rho^{(2)}(t)\bm{\hat{J}}^{(0)}(t)].
\end{equation}

Expanding the above equations with eigenbasis, the second order DC contributions  can be obtained as
\begin{eqnarray}
	j^{i}(0;\omega,-\omega)&&= j_a^{i}(0;\omega,-\omega)+ j_b^{i}(0;\omega,-\omega),\\
	j_a^{i}(0;\omega,-\omega)&&=e^3\sum_{m,n,l,\lambda,\bm{k}}[\frac{(f_{m}(\bm{k}+\frac{\lambda}{2}\bm{q})-f_l(\bm{k}-\frac{\lambda}{2}\bm{q}))  v^{j}_{m\bm{k}+\frac{\lambda}{2}\bm{q},l\bm{k}-\frac{\lambda}{2}\bm{q}}v^{k}_{l\bm{k}-\frac{\lambda}{2}\bm{q},n\bm{k}+\frac{\lambda}{2}\bm{q}} v^i_{n\bm{k}+\frac{\lambda}{2}\bm{q},m\bm{k}+\frac{\lambda}{2}\bm{q}}A_{j}(\lambda \omega)A_{k}(-\lambda \omega)}{(\epsilon_{m\bm{k}+\frac{\lambda}{2}\bm{q}}-\epsilon_{n\bm{k}+\frac{\lambda}{2}\bm{q}}-2i\hbar \eta)(\epsilon_{m\bm{k}+\frac{\lambda}{2}\bm{q}}-\epsilon_{l\bm{k}-\frac{\lambda}{2}\bm{q}}-\lambda\hbar\omega-i\hbar\eta)},\nonumber\\
	&&-\frac{(f_{l}(\bm{k}+\frac{\lambda}{2}\bm{q})-f_n(\bm{k}-\frac{\lambda}{2}\bm{q}))  v^{j}_{l\bm{k}+\frac{\lambda}{2}\bm{q},n\bm{k}-\frac{\lambda}{2}\bm{q}}v^{k}_{m\bm{k}-\frac{\lambda}{2}\bm{q},l\bm{k}+\frac{\lambda}{2}\bm{q}} v^i_{n\bm{k}- \frac{\lambda}{2}\bm{q},m\bm{k}-\frac{\lambda}{2}\bm{q}}A_{j}(\lambda \omega)A_{k}(-\lambda \omega)}{(\epsilon_{m\bm{k}-\frac{\lambda}{2}\bm{q}}-\epsilon_{n\bm{k}-\frac{\lambda}{2}\bm{q}}-2i\hbar \eta)(\epsilon_{l\bm{k}+\frac{\lambda}{2}\bm{q}}-\epsilon_{n\bm{k}-\frac{\lambda}{2}\bm{q}}-\lambda\hbar\omega-i\hbar\eta)}\\
	j_b^{i}(0;\omega,-\omega)=&&e^3\sum_{m,n,\bm{k},\lambda}\frac{(f_n(\bm{k}+\frac{\lambda}{2} \bm{q})-f_m(\bm{k}-\frac{\lambda}{2}\bm{q}))  v^{ik}_{m\bm{k}-\frac{\lambda}{2}\bm{q},n\bm{k}+\frac{\lambda}{2}\bm{q}}v^{j}_{n\bm{k}+\frac{\lambda}{2} \bm{q}, m\bm{k}-\frac{\lambda}{2} \bm{q}}A_{j}(\lambda\omega)  A_{k}(-\lambda\omega)}{\epsilon_{n\bm{k}+\frac{\lambda}{2} \bm{q}}-\epsilon_{m\bm{k}-\frac{\lambda}{2}\bm{q}}-\lambda \hbar \omega -i\hbar \eta}\nonumber\\&&+   
	\sum_{m,n,\lambda,\bm{k}}(\frac{\partial f_{n\bm{k}}}{\partial \epsilon_{n\bm{k}}}\delta_{mn}+\frac{f_{mn}(\bm{k})(1-\delta_{mn})}{\epsilon_{mn}(\bm{\bm{k}})-2i\hbar\eta})  v_{m\bm{k},n\bm{k}}^{jk}v^i_{n\bm{k},m\bm{k}A_{j}(\lambda\omega)A_{k}(-\lambda\omega)}. \label{cur} 
\end{eqnarray}

Here, $\epsilon_{mn}(\bm{k})=\epsilon_{m\bm{k}}-\epsilon_{n\bm{k}}$, $f_{mn}(\bm{k})=f_{m\bm{k}}-f_{n\bm{k}}$, the index of $j,k$ is implicitly summed over. For the convenience of presentation in later sections,  we have separated the photocurrents into two terms: $j_a^{i}(0;\omega,-\omega)$, $j_b^{i}(0;\omega,-\omega) $, where the former does not contain the second order derivation with respect to Hamiltonian, i.e., $v^{jk}$. The term involves $v^{jk}_{mn}$ with $m\neq n$, which could be finite when the Hamiltonian includes higher-order momentum terms.   On the other hand, the second term in Eq.~\eqref{cur} represents a Fermi surface photocurrent that is not related to the photo-drag effects, which is not relevant to the discussions in this work.


\subsection{Photon drag shift currents}

The notable resonant photocurrents: shift currents and injection currents are
mostly included in $j^{i}_a(0;\omega,-\omega)$. In the following two sections, let us see how the formalism for the shift currents and injection currents are modified.
We consider  $m\neq n$  contribution $j^{i}_0$ and $m=n$   contribution  $j^{i}_1$.  As we would show the former would give the shift currents, while the latter would give rise to the injection currents.

After rotating the indices, it can be shown that
\begin{eqnarray}
	j_0^{i}(0;\omega,-\omega)&&= - i \frac{e^3}{\hbar}\sum_{m\neq n,l,\lambda,\bm{k}} \frac{(f_n(\bm{k}-\frac{\lambda}{2}\bm{q})-f_{l}(\bm{k}+\frac{\lambda}{2}\bm{q}))  v^{j}_{l\bm{k}+\frac{\lambda}{2}\bm{q},n\bm{k}-\frac{\lambda}{2}\bm{q}}}{(\epsilon_{l\bm{k}+\lambda  \frac{\bm{q}}{2}}-\epsilon_{n\bm{k}-\frac{\lambda}{2}\bm{q}}-\lambda\hbar\omega-i\hbar\eta)} [-v^{k}_{n\bm{k}-\frac{\lambda}{2}\bm{q},m\bm{k}+\frac{\lambda}{2}\bm{q}}r_{ml}^{i}(\bm{k}+\frac{\lambda}{2}\bm{q})\nonumber\\
	\nonumber &&+v^{k}_{m\bm{k}- \frac{\lambda}{2}\bm{q},l\bm{k}+\frac{\lambda}{2}\bm{q}} r_{nm}^{i}(\bm{k}-\frac{\lambda}{2}\bm{q}) ]A_{j}(\lambda \omega)A_{k}(-\lambda \omega).\nonumber\\
\end{eqnarray}
Here, for the sake of convenience, we have defined
\begin{equation}
	r_{nm}(\bm{k})=\braket{n\bm{k}|i\partial_{k}|m\bm{k}}=\frac{ \hbar v_{nm}(\bm{k})}{i(\epsilon_{n}(\bm{k})-\epsilon_{m}(\bm{k}))},
\end{equation}
which can be regarded as the interband representation of position operation \cite{Sipe1995}.
Using the sum rule first $\sum_{m}r^{i}_{nm}v^{k}_{ml}-v^{k}_{nm}r^{i}_{ml}=-i(v^k_{nl})_{;k_i} $ \cite{Sipe1995},  we find
\begin{eqnarray}
	&&j_0^{i}(0;\omega,-\omega)= -  \frac{e^3}{\hbar}\sum_{n,l,\lambda,\bm{k}} \frac{(f_n(\bm{k}-\frac{\lambda}{2}\bm{q})-f_{l}(\bm{k}+\frac{\lambda}{2}\bm{q}))  }{(\epsilon_{l\bm{k}+\frac{\lambda}{2}\bm{q}}-\epsilon_{n\bm{k}- \frac{\lambda}{2}\bm{q}}-\lambda\hbar\omega-i\hbar\eta)} \nonumber\\
	&& v^{j}_{l\bm{k}+\frac{\lambda}{2}\bm{q},n\bm{k}- \frac{\lambda}{2}\bm{q}} (v^{k}_{n\bm{k}-\frac{\lambda}{2}\bm{q}, l\bm{k}+\frac{\lambda}{2}\bm{q}})_{;k_i}A_{j}(\lambda \omega)A_{k}(-\lambda \omega). \label{Eq_23}
\end{eqnarray}

The Fermi surface  principal part reads
\begin{eqnarray}
	&&j_{0,fs}^{i}(0;\omega,-\omega)=   \frac{e^3}{\hbar}\sum_{n,m,\lambda,\bm{k}} P[\frac{(f_n(\bm{k}+\frac{\lambda}{2}\bm{q})-f_{m}(\bm{k}-\frac{\lambda}{2}\bm{q}))  }{(\epsilon_{n\bm{k}+\frac{\lambda}{2}\bm{q}}-\epsilon_{m\bm{k}- \frac{\lambda}{2}\bm{q}}-\lambda\hbar\omega}]\nonumber\\
	&&v^{j}_{n\bm{k}+\frac{\lambda}{2}\bm{q},m\bm{k}- \frac{\lambda}{2}\bm{q}} (v^{k}_{m\bm{k}-\frac{\lambda}{2}\bm{q}, n\bm{k}+\frac{\lambda}{2}\bm{q}})_{;k_i}A_{j}(\lambda \omega)A_{k}(-\lambda \omega).
\end{eqnarray}
Note the index $(n,l)$ has been exchanged as $(m,n)$ here. After summing over $\lambda=\pm$, we find
\begin{eqnarray}
	&&j_{0,fs}^{i}(0;\omega,-\omega)=\frac{e^3}{\hbar} \sum_{m,n,\bm{k}} P[\frac{(f_n(\bm{k}+\frac{\bm{q}}{2})-f_{m}(\bm{k}-\frac{\bm{q}}{2}))}{\epsilon_{n\bm{k}+\frac{\bm{q}}{2}}-\epsilon_{m\bm{k}- \frac{\bm{q}}{2}}-\hbar\omega}]\nonumber\\
	&&\partial_{k_i}[v^{j}_{n\bm{k}+\frac{\bm{q}}{2},m\bm{k}- \frac{\bm{q}}{2}} v^{k}_{m\bm{k}-\frac{\bm{q}}{2}, n\bm{k}+\frac{\bm{q}}{2}}]A_{j}( \omega)A_{k}(-\omega). 
\end{eqnarray}

The resonant part of Eq.~\eqref{Eq_23} gives the so-called shift currents. After summing over $\lambda=\pm$, we find
\begin{eqnarray}
	j_{shift}^{i}(0;\omega,-\omega)&&=-\frac{i\pi e^3}{\hbar^2\omega^2}\sum_{n,m} (f_m(\bm{k}-\frac{\bm{q}}{2})-f_{n}(\bm{k}+\frac{\bm{q}}{2})) (v^{j}_{n\bm{k}+\frac{\bm{q}}{2},m\bm{k}-\frac{\bm{q}}{2}} (v^{k}_{m\bm{k}-\frac{\bm{q}}{2}, n\bm{k}+\frac{\bm{q}}{2}})_{;k_i}-v^{k}_{m\bm{k}-\frac{\bm{q}}{2},n\bm{k}+\frac{\bm{q}}{2}} (v^{j}_{n\bm{k}+\frac{\bm{q}}{2},m\bm{k}- \frac{\bm{q}}{2}})_{;k_i})\nonumber\\
	&&\delta(\omega_{n\bm{k}+\frac{\bm{q}}{2}}-\omega_{m\bm{k}-\frac{\bm{q}}{2}}-\omega) E_{j}(\omega)E_{k}(-\omega).
\end{eqnarray}
Note that all possible index $j,k$ should be sum over.

We can define
\begin{equation}
	j_{shift}^{i}(0;\omega,-\omega)=\sigma_{shift}^{i;j,k} (0; \omega,-\omega) E_{j}( \omega) E_{k}(- \omega).
\end{equation}

The shift current conductivity tensor is 
\begin{eqnarray}
	\sigma_{shift}^{i;j,k}(0; \omega,- \omega)&&=-i\pi \frac{e^3}{\hbar^2}\sum_{n, m} (f_m(\bm{k}-\frac{\bm{q}}{2})-f_{n}(\bm{k}+\frac{\bm{q}}{2})) (r^{j}_{n\bm{k}+\frac{\bm{q}}{2},m\bm{k}-\frac{\bm{q}}{2}} (r^{k}_{m\bm{k}-\frac{\bm{q}}{2}, n\bm{k}+\frac{\bm{q}}{2}})_{;k_i}-r^{k}_{m\bm{k}-\frac{\bm{q}}{2},n\bm{k}+\frac{\bm{q}}{2}} (r^{j}_{n\bm{k}+\frac{\bm{q}}{2},m\bm{k}- \frac{\bm{q}}{2}})_{;k_i})\nonumber\\
	&&\times  \delta(\omega_{n\bm{k}+\frac{\bm{q}}{2}}-\omega_{m\bm{k}-\frac{\bm{q}}{2}}-\omega).
\end{eqnarray}
Here, we have used the delta function to convert the velocity matrix into the interband position operator.

The useful trick in calculating the co-derivative numerically is  to use \cite{Junyeong2020} 
\begin{equation}
	(v_{mn}^{k})_{;k_i}= \partial_{k_i} v_{mn}^{k}-i(r_{mm}^{i}-r_{nn}^{i})v_{mn}^{k}
\end{equation}

and 
\begin{equation}
	\partial_{k_i} v_{mn}^{k}=v_{mn}^{ki}+i(r_{mm}^{i}-r_{nn}^{i}) v_{mn}^{k}-ir_{mn}^{i}\Delta_{mn}^{k}+\sum_{p\neq m,n}(\frac{v_{mp}^{i}v^k_{pn}}{\omega_{mp}}-\frac{v^k_{mp}v_{pn}^i}{\omega_{pn}})
\end{equation}
where $\Delta_{mn}^{k}=v_{mm}^{i}-v_{nn}^{k}$. Here we have included $v_{mn}^{ki}$ to take into account the higher-order derivative. It is easy to show that it is equivalent to include the resonant contribution from the first term of $j_b^{i}(0;\omega,-\omega)$. So we will not double count the resonant photon drag terms from $j_b^{i}(0;\omega,-\omega)$ below.  The delta function in the numerical calculation:
\begin{equation}
	\delta(\omega_{nm}-\omega)=\frac{1}{\pi}\text{Im}[\frac{1}{\omega_{nm}-\omega-i\eta}].
\end{equation}
In this work, we would fix $\hbar \eta=1$ meV for all the plots.

\subsection{  Photon drag injections currents}

The injection currents arises from $j_{1}^{i}$:
\begin{eqnarray}
	j_{1}^{i}(0;\omega,-\omega)&&\approx \frac{ie^3}{2\hbar \eta}\sum_{n,l,\lambda,\bm{k}}[\frac{(f_{n}(\bm{k})-f_l(\bm{k}-\lambda\bm{q}))  v^{j}_{n\bm{k},l\bm{k}-\lambda\bm{q}}v^{k}_{l\bm{k}-\lambda\bm{q},n\bm{k}} A_{j}(\lambda \omega)A_{k}(-\lambda \omega)}{(\epsilon_{n\bm{k}}-\epsilon_{l\bm{k}-\lambda\bm{q}}-\lambda\hbar\omega-i\hbar\eta)}\nonumber\\
	&&-\frac{(f_{l}(\bm{k}+\lambda\bm{q})-f_n(\bm{k}))  v^{j}_{l\bm{k}+\lambda\bm{q},n\bm{k}}v^{k}_{n\bm{k},l\bm{k}+\lambda\bm{q}} A_{j}(\lambda \omega)A_{k}(-\lambda \omega)}{(\epsilon_{l\bm{k}+\lambda\bm{q}}-\epsilon_{n\bm{k}}-\lambda\hbar\omega-i\hbar\eta)}]v^i_{n\bm{k},n\bm{k}}. \label{Eq_s19}
\end{eqnarray}
By exchanging the indices and shift the momentum, it can be rewritten as 
\begin{eqnarray}
	j_{1}^{i}(0;\omega,-\omega)&&\approx  -\frac{ie^3}{2\hbar \eta} \sum_{n,m,\lambda,\bm{k}} \frac{(f_{n}(\bm{k}+\frac{\lambda}{2}\bm{q})-f_m(\bm{k}-\frac{\lambda}{2}\bm{q}))(v_{n\bm{k}+\frac{\lambda}{2}\bm{q},n\bm{k}+\frac{\lambda}{2}\bm{q}}^{i}-v_{m\bm{k}-\frac{\lambda}{2}\bm{q},m\bm{k}-\frac{\lambda}{2}\bm{q}}^{i})   }{(\lambda\hbar\omega+i\hbar\eta-(\epsilon_{n\bm{k}+\frac{\lambda}{2}\bm{q}}-\epsilon_{m\bm{k}-\frac{\lambda}{2}\bm{q}}))}\nonumber\\
	&& \times v^{j}_{n\bm{k}+ \frac{\lambda}{2}\bm{q},m\bm{k}-\frac{\lambda}{2}\bm{q}}v^{k}_{m\bm{k}-\lambda\frac{\bm{q}}{2},n\bm{k}+ \frac{\lambda}{2}\bm{q}} A_{j}(\lambda \omega)A_{k}(-\lambda \omega).
\end{eqnarray}

Note that $j_1^{i}(t)=j_1(0;\omega,-\omega)e^{2\eta t}$  so that $dj_{1}/dt$ is the increase ratio of injection currents. But the injection currents upon a time of  $\frac{1}{2\eta}$ would be saturated so that $j_1^{i}(t)=\frac{1}{2\eta} dj_{1}/dt$. 
The resonant part of $j_1^{i}(0;\omega, -\omega)$ gives the so-called injection currents:
\begin{eqnarray}
	j_{inj}^{i}(0;\omega,-\omega)&&\approx  -\frac{\pi e^3}{2\hbar^2 \eta} \sum_{n,m,\lambda,\bm{k}} (f_{n}(\bm{k}+\frac{\lambda}{2}\bm{q})-f_m(\bm{k}-\frac{\lambda}{2}\bm{q}))(v_{n\bm{k}+\frac{\lambda}{2}\bm{q},n\bm{k}+\frac{\lambda}{2}\bm{q}}^{i}-v_{m\bm{k}-\frac{\lambda}{2}\bm{q},m\bm{k}-\frac{\lambda}{2}\bm{q}}^{i})   \nonumber\\
	&& \times v^{j}_{n\bm{k}+ \frac{\lambda}{2}\bm{q},m\bm{k}-\frac{\lambda}{2}\bm{q}}v^{k}_{m\bm{k}-\lambda\frac{\bm{q}}{2},n\bm{k}+ \frac{\lambda}{2}\bm{q}}\delta(\omega_{n\bm{k}+\frac{\lambda}{2}\bm{q}}-\omega_{m\bm{k}-\frac{\lambda}{2}\bm{q}}-\lambda\omega) A_{j}(\lambda \omega)A_{k}(-\lambda \omega).
\end{eqnarray}
By summing over $\lambda$ and using the delta function to convert the velocity matrix into the interband position matrix, we can find
\begin{equation}
	j_{inj}^{i}(0;\omega,-\omega)=\sigma_{inj}^{i;j,k} (0; \omega,- \omega) E_{j}( \omega) E_{k}(- \omega).
\end{equation}
Here, the injection  current conductivity tensor  can also be expressed as
\begin{eqnarray}
	\sigma_{inj}^{i;j,k}(0; \omega,- \omega)&&=-\frac{\pi e^3}{\hbar^2 \omega^2\eta } \sum_{n,m,\bm{k}} (f_{n}(\bm{k}+\frac{\bm{q}}{2})-f_m(\bm{k}-\frac{\bm{q}}{2}))(v_{n\bm{k}+\frac{\bm{q}}{2},n\bm{k}+\frac{\bm{q}}{2}}^{i}-v_{m\bm{k}-\frac{\bm{q}}{2},m\bm{k}-\frac{\bm{q}}{2}}^{i})\nonumber\\
	&&\times v^{j}_{n\bm{k}+ \frac{\bm{q}}{2},m\bm{k}-\frac{\bm{q}}{2}}v^{k}_{m\bm{k}-\frac{\bm{q}}{2},n\bm{k}+ \frac{\bm{q}}{2}}\delta(\omega_{n\bm{k}+\frac{\bm{q}}{2}}-\omega_{m\bm{k}-\frac{\bm{q}}{2}}-\omega).
\end{eqnarray}

The Fermi surface contributions from $j_{1}(i;\omega,-\omega)$ can be expressed as
\begin{eqnarray}
	j_{1,fs}^{i}(0;\omega,-\omega)&&\approx   -\frac{ ie^3}{2\hbar \eta} \sum_{n,m,\lambda,\bm{k}} (f_{n}(\bm{k}+\frac{\lambda}{2}\bm{q})-f_m(\bm{k}-\frac{\lambda}{2}\bm{q}))(v_{n\bm{k}+\frac{\lambda}{2}\bm{q},n\bm{k}+\frac{\lambda}{2}\bm{q}}^{i}-v_{m\bm{k}-\frac{\lambda}{2}\bm{q},m\bm{k}-\frac{\lambda}{2}\bm{q}}^{i})   \nonumber\\
	&& \times v^{j}_{n\bm{k}+ \frac{\lambda}{2}\bm{q},m\bm{k}-\frac{\lambda}{2}\bm{q}}v^{k}_{m\bm{k}-\lambda\frac{\bm{q}}{2},n\bm{k}+ \frac{\lambda}{2}\bm{q}}P[\frac{1}{\lambda\hbar\omega+i\hbar\eta-(\epsilon_{n\bm{k}+\frac{\lambda}{2}\bm{q}}-\epsilon_{m\bm{k}-\frac{\lambda}{2}\bm{q}})}] A_{j}(\lambda \omega)A_{k}(-\lambda \omega).
\end{eqnarray}

We can sum over $\lambda=\pm$,
\begin{equation}
	j_{1,fs}^{i}(0;\omega,-\omega)= -\frac{e^3}{\hbar} \sum_{m,n,\bm{k}} (f_{n}(\bm{k}+\frac{\bm{q}}{2})-f_m(\bm{k}-\frac{\bm{q}}{2})) v^{j}_{n\bm{k}+\frac{\bm{q}}{2},m\bm{k}-\frac{\bm{q}}{2}}  v^{k}_{m\bm{k}-\frac{\bm{q}}{2},n\bm{k}+\frac{\bm{q}}{2}} \partial_{k_i}P[\frac{1}{\hbar\omega-(\epsilon_{n\bm{k}+\frac{\bm{q}}{2}}-\epsilon_{m\bm{k}-\frac{\bm{q}}{2}})}] A_{j}(\omega) A_{k}(-\omega).
\end{equation}

\subsection{Fermi surface induced resonant photocurrents}

Then the Fermi surface photocurrents from the principle parts  in the previous sections are
\begin{eqnarray}
	j^{i}_{p,fs}&&=j^{i}_{0,fs}+j^{i}_{1,fs}\nonumber\\
	&&=-\frac{e^3}{\hbar} \sum_{m,n,\bm{k}} \partial_{k_i}[f_n(\bm{k}+\frac{\bm{q}}{2})-f_{m}(\bm{k}-\frac{\bm{q}}{2})]  \frac{v^{j}_{n\bm{k}+\frac{\bm{q}}{2},m\bm{k}- \frac{\bm{q}}{2}} v^{k}_{m\bm{k}-\frac{\bm{q}}{2}, n\bm{k}+\frac{\bm{q}}{2}}}{\epsilon_{n\bm{k}+\frac{\bm{q}}{2}}-\epsilon_{m\bm{k}- \frac{\bm{q}}{2}}-\hbar\omega} A_{j}( \omega)A_{k}(-\omega).
\end{eqnarray}
In the main text, we have used the above form to represent the Fermi surface-induced photocurrent contributions. It is worth noting that there are actually similar Fermi surface contributions from the photocurrents given by $j_{b}^{i}(0;\omega,-\omega)$, which we neglected in this work for simplicity.

\section{Shift currents with photon-drag from Floquet formalism}
In this section, we present a derivation of shift current formalism using the Floquet method.  We consider a model with time-reversal and inversion symmetry:
\begin{equation}
	H_{0}(\bm{k})=\begin{pmatrix}
		h(\bm{k})&0\\
		0& h^*(-\bm{k}).
	\end{pmatrix}
\end{equation}
After shining the light, according to ref.~\cite{Morimoto2016}, we can define the Floquet-Bloch Hamiltonian as 
\begin{equation}
	H^{FB}(\bm{k})=\begin{pmatrix}
		h_{\alpha}^{FB}(\bm{k},\bm{q}) & 0\\
		0& 	h_{\beta}^{FB}(\bm{k},\bm{q})
	\end{pmatrix}
\end{equation}
where the basis is ($\Psi_{1\alpha}(\bm{k}-\bm{q}/2),\Psi_{2\alpha}(\bm{k}+\bm{q}/2),  \Psi_{1\beta}(\bm{k}-\bm{q}/2),  \Psi_{2\beta}(\bm{k}+\bm{q}/2) $)$^T$ with $\alpha$ as a good quantum number (such as spin), and

\begin{eqnarray}
	h_{\alpha}^{FB}(\bm{k},\bm{q})=\begin{pmatrix}
		\epsilon_{1}(\bm{k}-\frac{\bm{q}}{2})+\Omega  &&	-i \bm{A_0} v^{\alpha}_{12}(\bm{k},\bm{q})\\
		i \bm{A_0}v^{\alpha}_{21}(\bm{k},\bm{q})	&&  	\epsilon_{2}(\bm{k}+\frac{\bm{q}}{2})  
	\end{pmatrix}, 	h_{\beta}^{FB}(\bm{k},\bm{q})=\begin{pmatrix}
		(\epsilon_{1}(\bm{k}-\frac{\bm{q}}{2})+\Omega)  &&	-i \bm{A_0}v^{\beta}_{12}(\bm{k},\bm{q})\\
		i \bm{A_0}v^{\beta}_{21}(\bm{k},\bm{q})	&&  	\epsilon_{2}(\bm{k}+\frac{\bm{q}}{2})  
	\end{pmatrix}	
\end{eqnarray}
Here, $\bm{A}_0=\frac{\bm{E}}{\Omega}$, $\Omega$ is the photon frequency, 
\begin{equation}
	v_{12}^{\alpha}(\bm{k},\bm{q})=  \braket{
		\Psi_{1\alpha}(\bm{k}-\frac{\bm{q}}{2})| \frac{\partial h(\bm{k})}{\hbar\partial \bm{k}}|\Psi_{2\alpha}(\bm{k}+\frac{\bm{q}}{2})}.
\end{equation}

Time-reversal symmetry 
$\ket{\Psi_{1\beta}(-\bm{k}+\bm{q}/2)}= \sum_{\alpha'} U_{\beta\alpha'}\ket{\Psi_{1\alpha'}(\bm{k}-\bm{q}/2)}^*$. As a result, due to time-reversal symmetry:
\begin{equation}
	h_{\alpha}^{FB}(\bm{k},\bm{q})=h_{\beta}^{FB*}(-\bm{k},-\bm{q}),\  	v_{12}^{\alpha}(\bm{k},\bm{q})=    -v_{21}^{\beta}(-\bm{k},-\bm{q}).
\end{equation}

The Green's function 
\begin{equation}
	G_{\alpha}^{R/L}(\bm{k},\bm{q})=\begin{pmatrix}
		\omega\mp \frac{i\Gamma}{2}-	\epsilon_{1}(\bm{k}-\frac{\bm{q}}{2})-\Omega & i\bm{A}_0 v^{\alpha}_{12}(\bm{k},\bm{q})\\
		- i\bm{A}_0  v^{\alpha}_{21}(\bm{k},\bm{q})	&	\omega\mp \frac{i\Gamma}{2}-\epsilon_{2}(\bm{k}+\frac{\bm{q}}{2})
	\end{pmatrix}^{-1}.
\end{equation}

We consider the DC limit so $\omega\mapsto 0^{+}$. 
We can define 
\begin{eqnarray}
	g_0=\omega\mp\frac{i\Gamma}{2}-\epsilon_0, g_x^{\alpha}=i\bm{A}_0\frac{v^{\alpha}_{12}(\bm{k},\bm{q})-v^{\alpha}_{21}(\bm{k},\bm{q})}{2},\\
	g_y^{\alpha}=-\bm{A}_0\frac{(v_{21}^{\alpha}(\bm{k},\bm{q})+v_{12}^{\alpha}(\bm{k},\bm{q}))}{2}, g_z=\frac{\epsilon_2(\bm{k}+\frac{\bm{q}}{2})-\epsilon_1(\bm{k}-\frac{\bm{q}}{2})-\Omega}{2}
\end{eqnarray}
where $\epsilon_0=\frac{\epsilon_1(\bm{k}-\frac{\bm{q}}{2})+\Omega+\epsilon_2(\bm{k}+\frac{\bm{q}}{2})}{2}$ is the origin of the Floquet-Bloch band and can be approximately set to be zero for a semiconductor.

Then the retarded Greene's function
\begin{equation}
	G_{\alpha}^{R}(\bm{k},\bm{q})= (g_0+\bm{g}\cdot \bm{\sigma})^{-1}= \begin{pmatrix}
		g_0+g_z& g_x^{\alpha}-ig_y^{\alpha}\\
		g_x^{\alpha}+ig_y^{\alpha}&g_0-g_z
	\end{pmatrix}^{-1}=\frac{g_0-\bm{g}\cdot \bm{\sigma}}{g_0^2-g^2}
\end{equation}
and the advance Green's function is obtained by replacing $g_0$ to $g_0^*$.

At the low temperature, the self-energy  is \cite{Morimoto2016} 
\begin{equation}
	\Sigma^{<}=\begin{pmatrix}
		i\Gamma f(\omega-\hbar\Omega)&0\\
		0&	i\Gamma f(\omega)
	\end{pmatrix} \approx i\Gamma \begin{pmatrix}
		1&0\\
		0&0
	\end{pmatrix}
\end{equation}

The lesser Green's function becomes
\begin{eqnarray}
	G^{<}(\bm{k},\bm{q})&&=\int \frac{d \omega}{2\pi} G^{R}\Sigma^{<}G^{A}\\
	&&=\int \frac{d\omega}{2\pi} i\Gamma\frac{  g_0-\bm{g}\cdot \bm{\sigma}}{g_0^2-g^2} \begin{pmatrix}
		1&0\\
		0&0
	\end{pmatrix}\frac{g_0^*-\bm{g}\cdot \bm{\sigma}}{(g_0^*)^2-g^2},\\
	&&=\int \frac{d\omega}{2\pi}\frac{i\Gamma}{(g_0^2-g^2)((g_0^*)^2-g^2)}\begin{pmatrix}
		(g_0-g_z)(g_0^*-g_z)&(g_0-g_z)(-g_x^{\alpha}+ig_y^{\alpha})\\
		(-g_x^{\alpha}-ig_y^{\alpha})(g_0^*-g_z)&(g_x^{\alpha})^2+(g_y^{\alpha})^2
	\end{pmatrix}
\end{eqnarray}
Replacing $g_0$ with $\omega\mp\frac{i\Gamma}{2}-\epsilon_0$,
\begin{eqnarray}
	G^{<}(\bm{k},\bm{q})&&=\frac{2i}{4g^2+\Gamma^2}\begin{pmatrix}
		(g^2+g_z^2)+\frac{\Gamma^2}{2} &(-g_z-i\frac{\Gamma}{2})(-g_x^{\alpha}+ig_y^{\alpha})\\
		(-g_z+i\frac{\Gamma}{2})(-g_x^{\alpha}-ig_y^{\alpha}) 	&g^2-g_z^2
	\end{pmatrix}\nonumber\\
	&&=\frac{i}{2}\sigma_0+\frac{2i}{4g^2+\Gamma^2}(g_x^{\alpha}g_z+\frac{\Gamma}{2}g_y^{\alpha})\sigma_x+\frac{2i}{4g^2+\Gamma^2}(g_y^{\alpha}g_z-\frac{\Gamma}{2}g_x^{\alpha})\sigma_y+\frac{i(g_z^2+\frac{\Gamma^2}{4})}{2(g^2+\frac{\Gamma^2}{4})}\sigma_z.
\end{eqnarray}

The current operator of the Floquet-Bloch  Hamiltonian
\begin{equation}
	\tilde{v}^{\alpha}_{i}(\bm{k},\bm{q})=	\frac{\partial h_{\alpha}^{FB}(\bm{k},\bm{q})}{\partial \bm{k}}=\begin{pmatrix}
		v_{11}(\bm{k}-\frac{\bm{q}}{2})  &&	-i\bm{A}_0(\frac{\partial v^{\alpha}(\bm{k},\bm{q})}{\partial \bm{k}_i})_{12}\\
		i\bm{A}_0(\frac{\partial v^{\alpha}(\bm{k},\bm{q})}{\partial \bm{k}_i})_{21}	&&  	v_{22}(\bm{k}+\frac{\bm{q}}{2})  
	\end{pmatrix}=f_0+\bm{f}\cdot \bm{\sigma}
\end{equation} 
\begin{eqnarray}
	-i\text{Tr}[\tilde{v}G^{<}]&&=f_0+\frac{g_z^2+\frac{\Gamma^2}{4}}{g^2+\frac{\Gamma^2}{4}}+\frac{(f_xg_x^{\alpha}+f_yg_y^{\alpha})g_z}{g^2+\frac{\Gamma^2}{4}}+\frac{\frac{\Gamma}{2}(f_xg_y^{\alpha}-f_yg_x^{\alpha})}{g^2+\frac{\Gamma^2}{4}}
\end{eqnarray}

The DC current ($\omega\rightarrow 0^+$) is given by \cite{Morimoto2016}
\begin{eqnarray}
	J_{i}&&=-ie \sum_{\alpha } \int_{\bm{k}} \text{Tr}[\tilde{v}^{\alpha}_{i}(\bm{k},\bm{q})	G_{\alpha}^{<}(\bm{k},\bm{q})].
\end{eqnarray} 
Let us separate the current into three parts:
\begin{eqnarray}
	J_i&&=j_{1i}+j_{2i}+j_{3i}\\
	j_{1i}&&= e \sum_{\alpha} \int_{\bm{k}} \frac{\frac{\Gamma}{2}(f_xg_y^{\alpha}-f_yg_x^{\alpha})}{g^2+\frac{\Gamma^2}{4}}.\\
	j_{2i}=&& e\sum_{\alpha} \int_{\bm{k}} \frac{(f_xg_x^{\alpha}+f_yg_y^{\alpha})g_z}{g^2+\frac{\Gamma^2}{4}}.\\
	j_{3i}&&= e\sum_{\alpha} \int_{\bm{k}} f_0+\frac{(g_z^2+\frac{\Gamma^2}{4})f_z}{g^2+\frac{\Gamma^2}{4}}.
\end{eqnarray} 
Following ref.~\cite{Morimoto2016}, we can show that
the shift current is given by $j_{1i}$:
\begin{eqnarray}
	j_{shift}=j_{1i}&&=\sum_{\alpha} \int_{\bm{k}}-\frac{\frac{\Gamma}{2}}{g^2+\frac{\Gamma^2}{4}} e|A_0|^2\text{Im}[ v^{\alpha}_{21}(\bm{k},\bm{q})(v^{\alpha}_{12}(\bm{k},\bm{q}))_{;k_i}]\\
	&&=\sum_{\alpha} \int_{\bm{k}} -e|A_0|^2\text{Im}[\frac{1}{g-\frac{\Gamma}{2}i}] \text{Im}[ v^{\alpha}_{21}(\bm{k},\bm{q})(v_{12}^{\alpha}(\bm{k},\bm{q}))_{;k_i}].
\end{eqnarray}
where we have used  the identity 
$v_{12}^{\alpha}(\bm{k},\bm{q}))_{;k_i}=-i v^{\alpha}_{12}(\bm{k},\bm{q})) R^{\alpha,k_i}_{12}(\bm{k},\bm{q})$
with $R^{\alpha,k_i}_{12}(\bm{k},\bm{q})=r^{k}_{11}-r^{k}_{22}+i\partial_{k_i} \log(v_{12}^{\alpha}(\bm{k},\bm{q}))$.
In the limit of zero $\Gamma$ limit and weak field limit $g_z\gg g_x,g_y$,
\begin{eqnarray}
	j_{shift}&&  \approx e\sum_{\alpha} -\int_{\bm{k}} 2\pi|A_0|^2 \delta (\epsilon_2(\bm{k}+\frac{\bm{q}}{2})-\epsilon_1(\bm{k}-\frac{\bm{q}}{2})-\Omega) \text{Im}[ v^{\alpha}_{21}(\bm{k},\bm{q})(v^{\alpha}_{12}(\bm{k},\bm{q}))_{;k_i}].
\end{eqnarray}

The above formalism is consistent with the main text equation (1) in the linear polarized light case. 
\section{Gauge invariance of photon drag formulism with doubly degenerated bands}

In the degenerated band case, we need to generate the formula:
\begin{equation}
	\sigma_{shift}^{i;j,k}(0; \omega,- \omega)=-\frac{i\pi e^3}{\hbar^2\omega^2}\sum_{n_{\alpha},l_{\beta}} (f_{n_\alpha \bm{k}}-f_{l_{\beta \bm{k}+\bm{q}}}) (v^{j}_{l_{\beta}\bm{k}+\bm{q},n_{\alpha}\bm{k}} (v^{k}_{n_{\alpha}\bm{k}, l_{\beta}\bm{k}+\bm{q}})_{;k_i}-v^{k}_{n_{\alpha}\bm{k},l_{\beta}\bm{k}+\bm{q}} (v^{j}_{l_{\beta}\bm{k}+\bm{q},n_{\alpha}\bm{k}})_{;k_i}) \delta(\omega_{l_{\beta}\bm{k}+\bm{q}}-\omega_{n_{\alpha}\bm{k}}-\omega). \label{Eq_105}
\end{equation}
Here, $\alpha=1, 2,.. M_{n}, \beta=1, 2,.. M_{l}$. Here, $M_l$ and $M_d$ are the dimensions of the manifold of $n-$the and $l-$ dimension. The injection and Fermi surface photon drag photocurrents can be generalized in a similar way. The co-derivative is generalized as 
\begin{equation}
	(v^{j}_{\tilde{n}\alpha_n,\tilde{m}\alpha_m})_{;k_i}= \partial_{k_i} v^{j}_{\tilde{n}\alpha_n,\tilde{m}\alpha_m}-i\sum_{\alpha} (r^{i}_{\tilde{n}\alpha_n,\tilde{n}\alpha}v^{j}_{\tilde{n}\alpha, \tilde{m}\alpha_m}-v^{j}_{\tilde{n}\alpha_n,\tilde{m}\alpha}r^{i}_{\tilde{m}\alpha, \tilde{m}\alpha_{m}}).
\end{equation}
The most physical scenario is the doubly degenerate case, where $M_{l}=M_{d}=2$ due to the presence of Kramers degeneracy.  The question is how the formula is invariant when a $U(2)$ gauge transformation is performed within a two-dimensional band manifold.  We can show that
\begin{eqnarray}
	\sum_{\alpha_1,\alpha_2}v^{j}_{m_{\alpha_2},n_{\alpha_1}}	v_{n_{\alpha_1},m_{\alpha_2}}^{k}=\text{Tr} [v^{j} v^{k}]= \text{Tr} [U_{m}^{\dagger}v^{j}U_{n}^{\dagger} U_{n} v^{k}U_{m}]
\end{eqnarray}
and
\begin{eqnarray}
	\sum_{\alpha_1,\alpha_2}v^{j}_{m_{\alpha_2},n_{\alpha_1}}	(v_{n_{\alpha_1},m_{\alpha_2}}^{k})_{;k_i}&&\mapsto \text{Tr}[ U_{m}^{\dagger} v^{j} U_{n} \{\partial_{k_i} (U_{n}^{\dagger} v^{k} U_{m})-i((	 U_{n}^{\dagger}r^{i}U_{n}+ iU_{n}^{\dagger}\partial_{k_i}U_{n}) U_{n}^{\dagger}v^{k}U_{m},\nonumber\\
	&&-U_{n}^{\dagger}v^{k}U_{m}(U_{m}^{\dagger}r^{i}U_{m}+ iU_{m}^{\dagger}\partial_{k_i}U_{m}))\}],\nonumber\\
	&&=\text{Tr}[v^{j}(\partial_{k_i}v^{k})-iv^{j}(r^{i}v^{k}-v^{k}r^{i})]+\text{Tr}[U_{m}^{\dagger}v^{j}U_{n}((\partial_{k_i}U_{n}^{\dagger})v^{k}U_{m}+U_n^{\dagger}(\partial_{k_i} U_{n}) U_{n}^{\dagger} v^{k}U_{m})]+\nonumber\\
	&&\text{Tr}[U_{m}^{\dagger}v^{j}U_{n}(U_{n}^{\dagger}v^{k}(\partial_{k_i}U_{m})-U_n^{\dagger} v^{k}(\partial_{k_i}U_{m}))],\nonumber\\
	&&=\text{Tr}[v^{j}(\partial_{k_i}v^{k})-iv^{j}(r^{i}v^{k}-v^{k}r^{i})]
\end{eqnarray}
Here, we have used the identify  $\partial_{k_i}(U^{\dagger} U)=0$. Because the energy is gauge invariant, we can see that the photon drag photocurrent formula is $U(2)$ gauge invariant. 
\section{Quantum geometry nature of photon drag BPVE}
The total photocurrents that include the shift currents and injections, and Fermi surface photocurrents:
\begin{eqnarray}
	j^{i}=(\sigma^{i;j,k}_{shift}+\sigma^{i;j,k}_{inj}+\sigma^{i;j,k}_{fs}) E_{j}(\omega) E_k(-\omega).
\end{eqnarray}
Here,
\begin{eqnarray}
	\sigma^{i;j,k}_{shift}&&=\frac{i\pi e^3}{\hbar^2\omega^2}\sum_{n,m,\bm{k}} (f_n(\bm{k}+\frac{\bm{q}}{2})-f_{m}(\bm{k}-\frac{\bm{q}}{2})) (v^{j}_{n\bm{k}+\frac{\bm{q}}{2},m\bm{k}-\frac{\bm{q}}{2}} (v^{k}_{m\bm{k}-\frac{\bm{q}}{2}, n\bm{k}+\frac{\bm{q}}{2}})_{;k_i}-v^{k}_{m\bm{k}-\frac{\bm{q}}{2},n\bm{k}+\frac{\bm{q}}{2}} (v^{j}_{n\bm{k}+\frac{\bm{q}}{2},m\bm{k}- \frac{\bm{q}}{2}})_{;k_i})\nonumber\\
	&&\delta(\omega_{n\bm{k}+\frac{\bm{q}}{2}}-\omega_{m\bm{k}-\frac{\bm{q}}{2}}-\omega).\\
	\sigma^{i;j,k}_{inj}&& =-\frac{\pi e^3}{\hbar^2 \omega^2\eta } \sum_{n,m,\bm{k}} (f_{n}(\bm{k}+\frac{\bm{q}}{2})-f_m(\bm{k}-\frac{\bm{q}}{2}))(v_{n\bm{k}+\frac{\bm{q}}{2},n\bm{k}+\frac{\bm{q}}{2}}^{i}-v_{m\bm{k}-\frac{\bm{q}}{2},m\bm{k}-\frac{\bm{q}}{2}}^{i}).\nonumber\\
	&&\times v^{j}_{n\bm{k}+ \frac{\bm{q}}{2},m\bm{k}-\frac{\bm{q}}{2}}v^{k}_{m\bm{k}-\frac{\bm{q}}{2},n\bm{k}+ \frac{\bm{q}}{2}}\delta(\omega_{n\bm{k}+\frac{\bm{q}}{2}}-\omega_{m\bm{k}-\frac{\bm{q}}{2}}-\omega).\\
	\sigma^{i;j,k}_{fs}&&=-\frac{e^3}{\hbar \omega^2} \sum_{m,n,\bm{k}} \partial_{k_i}[f_n(\bm{k}+\frac{\bm{q}}{2})-f_{m}(\bm{k}-\frac{\bm{q}}{2})]  \frac{v^{j}_{n\bm{k}+\frac{\bm{q}}{2},m\bm{k}- \frac{\bm{q}}{2}} v^{k}_{m\bm{k}-\frac{\bm{q}}{2}, n\bm{k}+\frac{\bm{q}}{2}}}{\epsilon_{n\bm{k}+\frac{\bm{q}}{2}}-\epsilon_{m\bm{k}- \frac{\bm{q}}{2}}-\hbar\omega}. \label{Eq_non2}
\end{eqnarray}

Let us expand the above formula up to the linear order in momentum $\bm{q}$. 

(i) Fermi distribution
\begin{eqnarray}
	f_n(\bm{k}+\frac{\bm{q}}{2})-f_m(\bm{k}-\frac{\bm{q}}{2})=f_{nm}(1-\delta_{mn})+q^{\lambda}v^{\lambda}_{nn}\frac{\partial f_n}{\partial \epsilon_n} \delta _{mn}
\end{eqnarray}
where $f_{nm}(\bm{k})=f_n(\bm{k})-f_m(\bm{k})$.

(ii) the interband velocity matrix:
using $\ket{n\bm{k}+\frac{\bm{q}}{2}}\approx \ket{n\bm{k}}-\frac{i}{2}\bm{q}\cdot \bm{r}_{nn}\ket{n\bm{k}}-\frac{i}{2}\sum_{l\neq n} \ket{l \bm{k}} \bm{q}\cdot \bm{r}_{ln}(\bm{k})$, we find
\begin{eqnarray}
	v^{j}_{m\bm{k}-\frac{\bm{q}}{2}, n\bm{k}+\frac{\bm{q}}{2}}&&\approx v^{j}_{mn}-iq^{\lambda}\Gamma^{\lambda j}_{mn}
\end{eqnarray}

where we have defined
\begin{equation}
	\Gamma^{\lambda j}_{mn}=\frac{1}{2}\{r^{\lambda}, v^{j}\}_{mn}.
\end{equation}

Similarly,  
\begin{equation}
	v^{j}_{n\bm{k}+\frac{\bm{q}}{2}, n\bm{k}+\frac{\bm{q}}{2}}\approx v^{j}_{nn}+\frac{q^{\lambda}}{2}\partial_{\lambda} v^{j}_{nn}.
\end{equation}

(iii) the derivative of the velocity matrix:
\begin{eqnarray}
	(v^{j}_{m\bm{k}-\frac{\bm{q}}{2}, n\bm{k}+\frac{\bm{q}}{2}})_{;k_i}&&= \partial_{k_i} v_{mn}^{j}-iq^{\lambda}\partial_{k_i}\Gamma_{mn}^{\lambda j} -i(r^{i}_{mm}(\bm{k}-\frac{\bm{q}}{2})-r^{i}_{nn}(\bm{k}+\frac{\bm{q}}{2})) (v_{mn}^{j}-iq^{\lambda}\Gamma_{mn}^{\lambda j})\nonumber\\
	&&\approx (v_{mn}^{j})_{;k_i}-iq^{\lambda} (\Gamma_{mn}^{\lambda j})_{;k_i}+i\frac{q^{\lambda}}{2}\partial_{k_\lambda}\Pi_{mn}^{i} v_{mn}^{j},
\end{eqnarray}
where $\Pi_{mn}^{i}=r_{mm}^{i}+r_{nn}^{i}$.

(iv) The energy difference
\begin{equation}
	\frac{1}{\epsilon_{n\bm{k}+\frac{\bm{q}}{2}}-\epsilon_{m\bm{k}- \frac{\bm{q}}{2}}-\hbar\omega-i\hbar \eta}\approx\frac{1}{\epsilon_{n\bm{k}}-\epsilon_{m\bm{k}}-\hbar\omega-i\hbar\eta}-\frac{\frac{q^{\lambda}}{2} (v^{\lambda}_{nn}+v^{\lambda}_{mm})}{(\epsilon_{n\bm{k}}-\epsilon_{m\bm{k}}-\hbar\omega-i\hbar \eta)^2};
\end{equation}
\begin{equation}
	\delta(\omega_{n\bm{k}+\frac{\bm{q}}{2}}-\omega_{m\bm{k}-\frac{\bm{q}}{2}}-\omega)\approx \delta (\omega_{n\bm{k}}-\omega_{m\bm{k}}+\frac{q^\lambda}{2}(\partial_{\lambda}\omega_{n\bm{k}}+\partial_{\lambda}\omega_{m\bm{k}}))-\omega).
\end{equation}

Then let us we expand
\begin{equation}
	\sigma^{i;j,k}(\bm{q})\approx \sigma^{i;j,k}(\bm{q}=0)+q_{\lambda}\sigma^{\lambda ijk}.
\end{equation}

In the following, we shall focus on the interband contributions to the photocurrents. The leading order inter-band photon-drag induced contributions are
\begin{eqnarray}
	\sigma^{\lambda ijk}_{shift}
	&&\approx \frac{\pi e^3 }{\hbar^2\omega^2} \sum_{m\neq n,\bm{k}} f_{nm}[(v_{nm}^{j}(\Gamma^{\lambda k}_{mn})_{;k_i}+v_{mn}^{k}(\Gamma_{nm}^{\lambda j})_{;k_i})-(\Gamma^{\lambda j}_{nm}(v_{mn}^{k})_{;k_i}+\Gamma^{\lambda k}_{mn}(v_{nm}^{j})_{;k_i})]\delta (\omega_{nm}-\omega), \nonumber \\
	&& -\frac{\pi e^3 }{\hbar^2\omega^2} \sum_{m\neq n,\bm{k}} f_{nm}v_{nm}^{j}v_{mn}^{k}\partial_{k_\lambda}\Pi_{mn}^{i} \delta(\omega_{nm}-\omega) \\ 
	\sigma^{\lambda ijk}_{inj}
	&&\approx -\frac{\pi e^3 }{\hbar^2 \omega^2\eta } \sum_{m\neq n,\bm{k}} f_{nm} [i\Delta_{nm}^{i} (\Gamma_{nm}^{\lambda j}v_{mn}^{k}-v_{nm}^{j}\Gamma^{\lambda k}_{mn})+\frac{1}{2}\partial_{\lambda}\Delta_{nm}^{i} v_{nm}^{j} v_{mn}^{k}]\delta(\omega_{nm}-\omega),\\
	\sigma^{\lambda ijk}_{fs} 
	&&\approx -\frac{ i e^3 q^{\lambda}}{\hbar \omega^2} \sum_{m\neq n,\bm{k}} \partial_{k_i}f_{nm} \frac{ \Gamma_{nm}^{\lambda j} v_{mn}^{k}-v_{nm}^{j}\Gamma^{\lambda k}_{mn}}{\epsilon_{nm}-\hbar \omega} 
\end{eqnarray}
where $\Delta_{nm}^{i}=v_{nn}^{i}-v^{i}_{mm}$.

To see the geometry meaning of the above inter-band terms, we can rewrite
\begin{eqnarray}
\Gamma_{mn}^{\lambda j}&&=\frac{1}{2}\sum_{l} (r^{\lambda}_{m l} v^{j}_{ln}+ v^{j}_{ml}r^{\lambda}_{l n}),\nonumber\\
&&=\frac{1}{2}\sum_{l\neq m,n} (r_{ml}^{\lambda}v_{ln}^{j}+v^{j}_{ml}r^{\lambda}_{l n})+\frac{1}{2} (r_{mm}^{\lambda}v_{mn}^{j}+v^{j}_{mm}r^{\lambda}_{mn})+\frac{1}{2} (r_{mn}^{\lambda}v_{nn}^{j}+v^{j}_{mn}r^{\lambda}_{n n}),\nonumber\\
&&=\frac{1}{2}\sum_{l\neq m,n} (r_{ml}^{\lambda}v_{ln}^{j}+v^{j}_{ml}r^{\lambda}_{l n})+\frac{1}{2} (r_{mm}^{\lambda}+r_{nn}^{\lambda})v^{j}_{mn}+\frac{1}{2} r^{\lambda}_{mn}(v_{mm}^{j}+v_{nn}^{j}),
\end{eqnarray}
where $r_{nm}=\frac{ v_{nm}}{i\omega_{nm}}$. In general, there are two-band and three-band contributions in $\Gamma_{mn}^{\lambda j}$. For simplicity, let us focus on two band contributions 
\begin{equation}
\Gamma_{mn,2}^{\lambda j}=\frac{1}{2} r_{mn}^{\lambda}(v_{nn}^{j}+v_{mm}^{j})+\frac{1}{2} (r_{mm}^{\lambda}+r_{nn}^{\lambda})v^{j}_{mn}=\frac{1}{2}r_{mn}^{\lambda} W_{nm}^{j}+\frac{1}{2} \Pi_{mn}^{\lambda}v_{mn}^{j}.
\end{equation}

The optical transitions usually happen between  conductance and valence bands. The first term is determined by the particle-hole band velocity asymmetry, while the second term is determined by the particle-hole Berry connection asymmetry.

\emph{ Particle-hole velocity asymmetry term}:  In this limit, 
\begin{equation}
\Gamma_{mn,2}^{\lambda j}\approx \frac{1}{2} r_{mn}^{\lambda}(v_{nn}^{j}+v_{mm}^{j})
\end{equation}

The leading order inter-band  contributions from the two-band process are given by
\begin{eqnarray}
\sigma^{\lambda ijk}_{shift,2}&&\approx \frac{i\pi e^3 }{2 \hbar^2\omega}  \sum_{m\neq n,\bm{k}} f_{nm} [(v_{nn}^{k}+v_{mm}^{k})(r_{nm}^{j}(r^{\lambda}_{mn})_{;k_i}-r^{\lambda}_{mn}(r_{nm}^{j})_{;k_i})\nonumber\\
&&+(v_{nn}^{j}+v_{mm}^{j})(r_{nm}^{\lambda}(r^{k}_{mn})_{;k_i}- r_{mn}^{k}(r_{nm}^{\lambda})_{;k_i})+(\partial_{k_i}v_{nn}^{k}+\partial_{k_i}v_{mm}^{k})r_{nm}^{j} r_{mn}^{\lambda}\nonumber\\
&&-r^{k}_{mn}r_{nm}^{\lambda}(\partial_{k_i}v_{nn}^{j}+\partial_{k_i}v_{mm}^{j})]\delta(\omega_{nm}-\omega).
\end{eqnarray}
It is worth noting that
\begin{equation}
(r^j_{mn}(r^k_{nm})_{;k_i}-(r^j_{mn})_{;k_i}r^k_{nm})=i(R^{i;j}_{mn}-R^{i;k}_{nm}) r^{j}_{mn}r_{nm}^k,
\end{equation}
where the shift vector  $R_{nm}^{i;j}=\xi^i_{nn}-\xi_{mm}^{i}+i\partial_{k_i}\log(r_{nm}^j)$. 
Let us also define the geometric tensor
\begin{equation}
G_{mn}^{jk}=r_{mn}^{j}r_{nm}^{k}=\mathcal{G}_{mn}^{jk}-\frac{i}{2}\Omega_{mn}^{jk},
\end{equation}
where the quantum metric and Berry curvature tensor are given by 
\begin{eqnarray}
\mathcal{G}_{mn}^{jk}=\frac{1}{2}(r^{j}_{mn}r_{nm}^{k}+r_{mn}^{k}r_{nm}^j)=\text{Re}[r^j_{mn}r^k_{nm}],\\
\Omega_{mn}^{jk}=i(r_{mn}^{j}r_{nm}^k-r_{mn}^kr_{nm}^j)=-2\text{Im}[r^j_{mn}r^k_{nm}].
\end{eqnarray}

We find
\begin{eqnarray}
\sigma^{\lambda ijk}_{shift, 2}&&\approx  \frac{\pi e^3 }{2 \hbar^2\omega} \sum_{m\neq n, \bm{k}} f_{nm}W_{nm}^{k}[(R^{i;\lambda}_{mn}-R^{i;j}_{nm})G^{\lambda j}_{mn}+i\partial_{k_i} G_{mn}^{\lambda 
j}]\delta(\omega_{nm}-\omega)+\\
&&f_{nm}W_{nm}^{j}[(R_{mn}^{i;\lambda}-R_{nm}^{i;k})G^{k\lambda}_{mn}-i\partial_{k_i}G^{k\lambda}_{mn}]\delta(\omega_{nm}-\omega).
\end{eqnarray}

In the linearly polarized light with $j=k$, we can simplify
\begin{equation}
\sigma^{\lambda ijj}_{shift, 2}\approx \frac{\pi e^3 }{ \hbar^2\omega} \sum_{m\neq n, \bm{k}} f_{nm}W_{nm}^{j}[(R_{mn}^{i;\lambda}-R_{nm}^{i;j})\mathcal{G}_{mn}^{\lambda j}+\frac{1}{2}\partial_{k_i}\Omega_{mn}^{\lambda j}]\delta(\omega_{nm}-\omega).
\end{equation}
When $j\neq k$, the imaginary part of $\sigma_{shift, inter,2}^{\lambda i j k}$ would be finite:
\begin{eqnarray}
\text{Im}[ \sigma^{\lambda ijk}_{shift,  2}]&&=\frac{\pi e^3}{4 \hbar^2\omega} \sum_{m\neq n, \bm{k}} f_{nm}W_{nm}^{k} [-(R^{i;\lambda}_{mn}-R^{i;j}_{nm})\Omega^{\lambda j}_{mn}+\partial_{k_i} \mathcal{G}_{mn}^{\lambda 
j}]\delta(\omega_{nm}-\omega)+\nonumber\\
&&f_{nm}W_{nm}^{j}[-(R_{mn}^{i;\lambda}-R_{nm}^{i;k})\Omega^{k\lambda}_{mn}-i\partial_{k_i}\mathcal{G}^{k\lambda}_{mn}]\delta(\omega_{nm}-\omega).
\end{eqnarray}
This corresponds to a circular shift effect. 

Similarly, we can find the injection current from the two-band process is given by
\begin{eqnarray}
\sigma^{\lambda ijk}_{inj, 2} &&\approx -\frac{\pi e^3 }{\hbar^2 \omega^2\eta } \sum_{m\neq n,\bm{k}} f_{nm} [i\Delta_{nm}^{i} (\Gamma_{nm}^{\lambda j}v_{mn}^{k}-v_{nm}^{j}\Gamma^{\lambda k}_{mn})+\frac{1}{2}\partial_{\lambda}\Delta_{nm}^{i} v_{nm}^{j} v_{mn}^{k}]\delta(\omega_{nm}-\omega)\nonumber\\
&&\approx -\frac{\pi e^3 }{\hbar^2 \eta } \sum_{m\neq n,\bm{k}} f_{nm} [\frac{1}{2\omega}\Delta_{nm}^{i} ((v_{nn}^{j}+v_{mm}^{j})G_{mn}^{k\lambda}+(v_{nn}^{k}+v_{mm}^{k})G_{mn}^{\lambda j}-\frac{1}{2}\Delta_{nm}^{i} \partial_{k_\lambda} G_{mn}^{kj}]\delta(\omega_{nm}-\omega) \nonumber
\end{eqnarray}
Hence, for linearly polarized light with $j=k$,
\begin{eqnarray}
\sigma^{\lambda ijj}_{inj, 2}=-\frac{\pi e^3 }{\hbar^2 \eta } \sum_{m\neq n,\bm{k}}f_{nm}[\frac{1}{\omega}\Delta_{nm}^{i} (W_{nm}^{j}\mathcal{G}^{j\lambda}_{mn}- \frac{1}{2}\Delta_{nm}^{i} \partial_{k_\lambda} (\mathcal{G}_{mn}^{jj})]\delta(\omega_{nm}-\omega) .
\end{eqnarray}
When $j\neq k$, 
\begin{equation}
\text{Im} [\sigma^{\lambda ijk}_{inj, 2}]=\frac{\pi e^3 }{4\hbar^2 \eta } \sum_{m\neq n,\bm{k}} f_{nm} [\frac{1}{\omega}\Delta_{nm}^{i} (W_{nm}^{j}\Omega_{mn}^{k\lambda}+W_{nm}^{k}\Omega_{mn}^{\lambda j}-\Delta_{nm}^{i} \partial_{k_\lambda} \Omega_{mn}^{kj}]\delta(\omega_{nm}-\omega).
\end{equation}
The Fermi surface induced inter-band contributions are given by
\begin{eqnarray}
\sigma^{\lambda ijk}_{fs,2} &&\approx -\frac{ i e^3 }{\hbar \omega^2} \sum_{m\neq n,\bm{k}} \partial_{k_i}f_{nm} \frac{ \Gamma_{nm}^{\lambda j} v_{mn}^{k}-v_{nm}^{j}\Gamma^{\lambda k}_{mn}}{\epsilon_{nm}-\hbar \omega} \nonumber\\
&&=\frac{e^3 }{2\hbar^2 \omega^2} \sum_{m\neq n,\bm{k}}(\partial_{k_i} f_{nm})[-r_{nm}^{\lambda}r_{mn}^{k}(v_{nn}^j+v_{mm}^{j})-r_{nm}^{j}r_{mn}^{\lambda}(v_{nn}^{k}+v_{mm}^{k})](1+\frac{\hbar \omega}{\epsilon_{nm}-\hbar\omega}),\nonumber\\
&&=\frac{e^3 }{2\hbar^2 \omega^2} \sum_{m\neq n,\bm{k}} f_{nm}[\partial_{k_i} (G_{mn}^{k\lambda})\frac{W_{nm}^{j}\epsilon_{nm}}{\epsilon_{nm}-\hbar\omega}+G_{mn}^{k\lambda}\partial_{k_i}(\frac{W_{nm}^{j}\epsilon_{nm}}{\epsilon_{nm}-\hbar\omega}) +(j\leftrightarrow k)^{*}]
\end{eqnarray}

For linear polarized light with $j=k$, we find
\begin{eqnarray}
\sigma^{\lambda ijj}_{fs,2}\approx \frac{e^3 }{\hbar^2 \omega^2} \sum_{m\neq n,\bm{k}} f_{nm} \{(\partial_{k_i}(g_{mn}^{j\lambda})W_{nm}^{j}\frac{\epsilon_{nm}}{\epsilon_{nm}-\hbar\omega}+g_{mn}^{j\lambda}\partial_{k_i}[\frac{\epsilon_{nm}W_{nm}^{j}}{\epsilon_{nm}-\hbar\omega}]\}
\end{eqnarray}
When $j\neq k$, the imaginary part is
\begin{eqnarray}
\text{Im}[\sigma^{\lambda ijk}_{fs,2}]&&=\frac{e^3 }{4\hbar^2 \omega^2} \sum_{m\neq n,\bm{k}}(\partial_{k_i} f_{nm})[\Omega_{mn}^{k\lambda}W_{nm}^{j}+\Omega_{mn}^{\lambda j}W_{nm}^{k}]\frac{\epsilon_{nm}}{\epsilon_{nm}-\hbar\omega}\\
&&=-\frac{e^3 }{4\hbar^2 \omega^2} \sum_{m\neq n,\bm{k}} f_{nm}\{(\partial_{k_i}\Omega_{mn}^{k\lambda})\frac{\epsilon_{nm}W_{nm}^{j}}{\epsilon_{nm}-\hbar\omega}+ \Omega_{mn}^{k\lambda}\partial_{k_i}[\frac{\epsilon_{nm}W_{nm}^{j}}{\epsilon_{nm}-\hbar\omega}]+(\partial_{k_i}\Omega_{mn}^{\lambda j})\frac{\epsilon_{nm}W_{nm}^{k}}{\epsilon_{nm}-\hbar\omega}+\nonumber\\
&&\Omega_{mn}^{\lambda j}\partial_{k_i}[\frac{\epsilon_{nm}W_{nm}^{k}}{\epsilon_{nm}-\hbar\omega}]\}.
\end{eqnarray}

\emph{ Particle-hole Berry connection asymmetry term}:

As an illustration, we now look at the particle-hole Berry connection asymmetry term for the shift current. 
\begin{eqnarray}
\sigma^{\lambda ijk}_{shift}
&&\approx \frac{\pi e^3 }{\hbar^2\omega^2} \sum_{m\neq n,\bm{k}} f_{nm}[(v_{nm}^{j}(\Gamma^{\lambda k}_{mn})_{;k_i}+v_{mn}^{k}(\Gamma_{nm}^{\lambda j})_{;k_i})-(\Gamma^{\lambda j}_{nm}(v_{mn}^{k})_{;k_i}+\Gamma^{\lambda k}_{mn}(v_{nm}^{j})_{;k_i})]\delta (\omega_{nm}-\omega)\nonumber\\
&&-\frac{\pi e^3 }{\hbar^2\omega^2} \sum_{m\neq n,\bm{k}} f_{nm}v_{nm}^{j}v_{mn}^{k}\partial_{k_\lambda}\Pi_{mn}^{i} \delta(\omega_{nm}-\omega)\nonumber\\
&&= \frac{\pi e^3 }{2\hbar^2\omega^2} \sum_{m\neq n,\bm{k}} f_{nm} [2v_{nm}^{j}v_{mn}^{k}\partial_{k_i}\Pi_{mn}^{\lambda}+ \Pi_{mn}^{\lambda}(v_{nm}^{j}(v_{mn}^{k})_{;k_i}+v_{mn}^{k}(v_{nm}^{j})_{;k_i})-\nonumber\\
&&\Pi_{mn}^{\lambda}(v_{nm}^{j}(v_{mn}^{k})_{;k_i}+v_{mn}^{k}(v_{nm}^{j})_{;k_i})]\delta (\omega_{nm}-\omega)-\frac{\pi e^3 }{\hbar^2\omega^2} \sum_{m\neq n,\bm{k}} f_{nm}v_{nm}^{j}v_{mn}^{k}\partial_{k_\lambda}\Pi_{mn}^{i} \delta(\omega_{nm}-\omega)\nonumber\\
&&=\frac{\pi e^3 }{\hbar^2\omega^2} \sum_{m\neq n,\bm{k}} f_{nm} v_{nm}^{j}v_{mn}^{k} (\partial_{k_i}\Pi_{mn}^{\lambda}-\partial_{k_\lambda}\Pi_{mn}^{i}) \delta (\omega_{nm}-\omega)\nonumber\\
&&=\frac{2\pi e^3 }{\hbar^2\omega^2} \sum_{m\neq n,\bm{k}} f_{nm} v_{nm}^{j}v_{mn}^{k} (\Omega_{m}^{i\lambda}+ \Omega_{n}^{i\lambda}) \delta (\omega_{nm}-\omega).
\end{eqnarray}
Interestingly, the term involves the summation of Berry curvature between two bands. In this work, the Berry curvature is typically opposite for particle and hole bands. For the simplification of presentations,   we have neglected this particle-hole Berry connection asymmetric term in the main text. However, in a more general case, this contribution can be finite.

\section{Polarization dependence of photocurrents analysis}
As shown in the main text Fig.1, the basis vectors of the light frame $\hat{x'}\hat{y'}\hat{z'}$ 
and the sample frame  $\hat{x}\hat{y}\hat{z}$  are related through $\hat{x'}=\hat{x},
\hat{y'}=\cos\theta \hat{y}-\sin\theta\hat{z},
\hat{z'}=\sin\theta \hat{y}+\cos\theta \hat{z}$.
The photon momentum direction is 
\begin{equation}
\bm{q}=q\hat{z'}=q\sin\theta\hat{y}+q\cos\theta \hat{z}.
\end{equation}
The conductivity tensor is
\begin{equation}
\sigma^{i;jk}(0;\omega,-\omega)=\sigma_{R}^{i;jk}+i\sigma_{I}^{i;jk}.
\end{equation}
Here, because $E(\omega)=E^*(-\omega)$, $\sigma_{R}^{i;jk}=\sigma_{R}^{i;kj}$, and $\sigma_{I}^{i;jk}=-\sigma_{I}^{i;kj}$. 
The polarization dependence of light is tuned by 
\begin{eqnarray}
\bm{E}(t)&&=|E|e^{i\bm{q}\cdot \bm{r}-i\omega t}[(\cos^2\alpha+i\sin^2\alpha)\hat{x'}+(\sin\alpha\cos\alpha (1-i)) \hat{y'}]+c.c..,\nonumber\\
&&=|E|e^{i\bm{q}\cdot \bm{r}-i\omega t}[(\cos^2\alpha+i\sin^2\alpha)\hat{x}+(\sin\alpha\cos\alpha (1-i))(\cos\theta \hat{y}-\sin\theta \hat{z})]+c.c..
\end{eqnarray}
As a result, the photocurrents at different polarization are captured by
\begin{eqnarray}
J_{\alpha}^{i}(0;\omega,-\omega)&&=[(\cos^4\alpha+\sin^4\alpha)\sigma_{R}^{i;xx}(\bm{q})\nonumber\\
&&+ \frac{1}{2}\cos^2\theta\sin^2(2\alpha)\sigma_{R}^{i;yy}(\bm{q})+\frac{\cos\theta}{2}\sin(4\alpha)\sigma_{R}^{i;xy}(\bm{q})-\cos\theta \sin(2\alpha)\sigma_{I}^{i;xy}(\bm{q})]|E|^2.
\end{eqnarray}

\section{Photon drag  BPVE with intra-band contributions in a 2D Dirac semimetal}

\begin{figure}[h]
\centering
\includegraphics[width=0.5\linewidth]{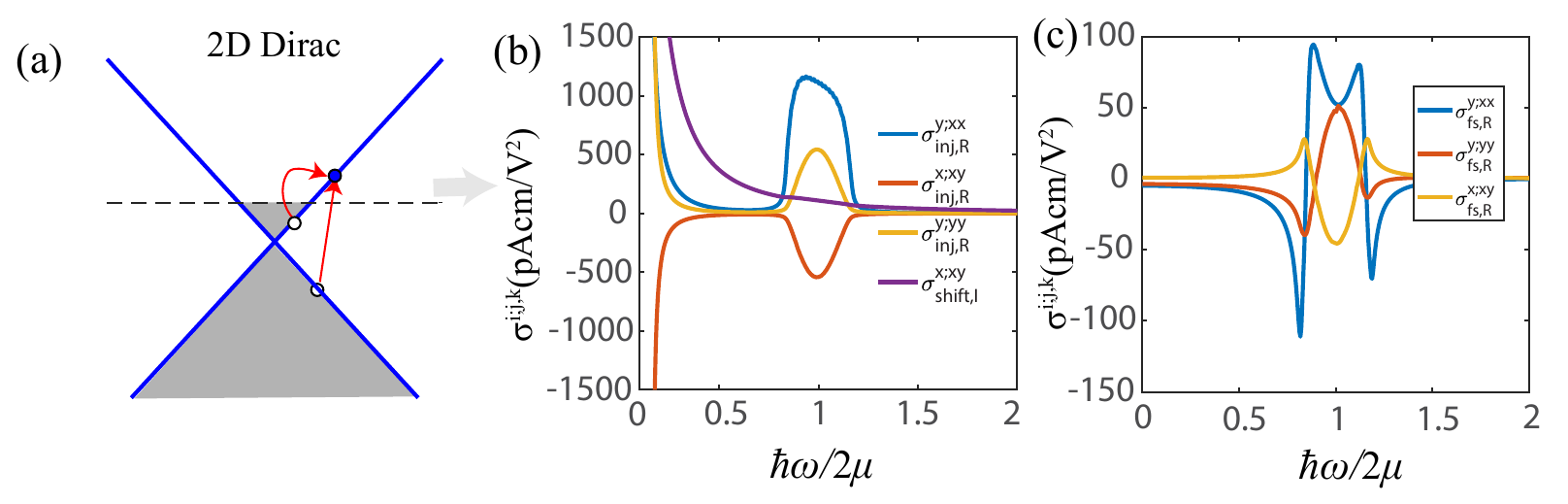}
\caption{The photon drag BPVE in 2D Dirac semimetal. (a) shows the intra and interband photoexcitations under photon drag in a Dirac semimetal. (b) and (c) show various non-zero shift/injection/Fermi surface-induced photoconductivity tensors from the photoexcitation of (a).  }
\label{fig:figs1}
\end{figure}
One may wonder what would happen for the photon drag photocurrents in a non-magnetic semimetal instead of an insulator. As an exploration, we now calculate the photocurrent conductivity tensor of a 2D Dirac Hamiltonian of graphene $H=v_F(k_xs_x+\tau k_ys_y)-\mu$ with $v_F=2.338$eV, $\tau=\pm$ labels the valley, $s_i$ are Pauli matrices in sublattice space. As shown in Fig.~\ref{fig:figs1}(a), unlike the insulating case,  there exhibits both intra-band and inter-band photoexcitations under photon drag.  The frequency dependence of all nonvanishing photocurrent conductivity tensors is shown in Figs.~\ref{fig:figs1}(b) and (c), while the polarization dependence is the same as the insulator case so we do not repeat it here. It can be seen that the intra-band excitations contribute to a diverging Drude-like peak at a low-frequency limit.  Moreover, as expected, the Fermi surface-induced photon-drag photocurrents become finite [Fig.~\ref{fig:figs1}(c)].

\section{Noncentrosymmetic shift photoconductivity tensor from WTe2 model}
\begin{figure}
\centering
\includegraphics[width=0.5\linewidth]{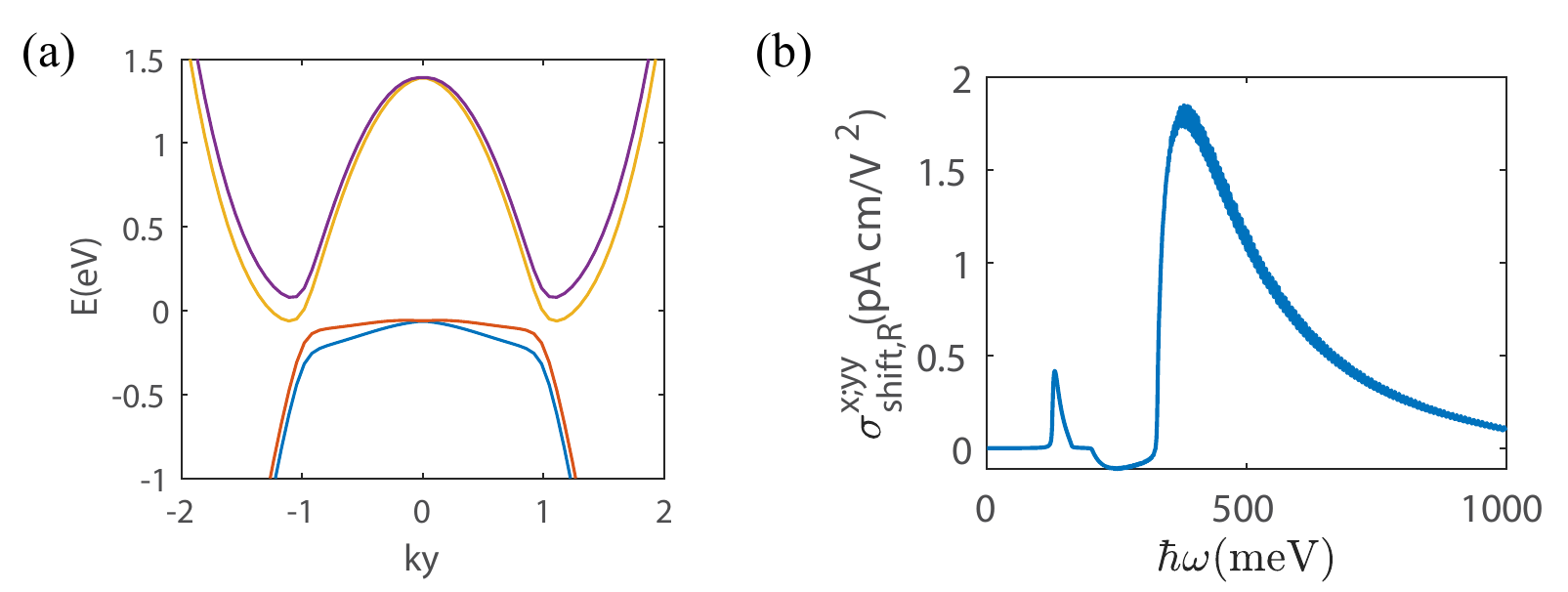}
\caption{(a) The band structure of WTe2 model without inversion symmetry. The inversion is broken through a Rashba spin-orbit coupling term. (b) The $\sigma_{shift, R}^{x;yy}$ versus the photon energy from the noncentrosymmetric WTe2 model }
\label{fig:figs2}
\end{figure}

The inversion symmetry of WTe2 model can be broken by taking into account of the gate effects, which generate an out-of-plane electric field. We can capture the effects of inversion symmetry breaking in this monolayer sample with a Rashba SOC term.  The resulting band structure is shown in Fig.~\ref{fig:figs2}(a).

The calculated photoconductivity tensor at zero-q for this noncentrosymmetric WTe2 model. As an illustration, $\sigma_{shift,R}^{x;yy}$ versus photon energy $\hbar\omega$ are plotted in Fig.~\ref{fig:figs2}(b).  Compared to the photon drag shift photocurrent [main text Fig.2], we can see that (i) The peak value of the photon drag shift photocurrent and noncentrosymmetric shift photocurrent are at the same order with a $q_y$=0.01 nm$^{-1}$. Note that the photon drag shift photocurrent can be reduced significantly if the photon momentum q is reduced.  (ii) The noncentrosymmetric shift photocurrents are sizable at a much wider frequency region. (iii)  As expected, the noncentrosymmetric shift photocurrent in WTe2 can be excited with linearly polarized light, while the photon drag photocurrent is excited with circularly polarized light.

\section{Fermi surface photocurrents and nonlinear transports}

In this section, we illustrate the connection between the Fermi surface photocurrents and nonlinear transports resulting from Berry curvature dipole and quantum metric.
The Fermi surface photocurrents are written as
\begin{equation}
j^{i}_{fs}=\frac{e^3}{\hbar \omega^2} \sum_{n,m} f_{nm} \partial_{k_i} (  \frac{v^{j}_{n,m} v^{k}_{m, n}}{\epsilon_{n}-\epsilon_{m}-\hbar\omega}) E_{j}(\omega)E_{k}(-\omega).
\end{equation}

We now focus on the interband contribution only ($m\neq n$).
Use the identity $\frac{1}{\epsilon_{nm}-\lambda\hbar\omega}=\frac{\lambda\hbar\omega}{\epsilon_{nm}(\epsilon_{nm}-\lambda\hbar\omega)}+ \frac{1}{\epsilon_{nm}}$, we can show
\begin{eqnarray}
\frac{e^3}{\omega}\sum_{m\neq n}f_{nm}\partial_{i}(\frac{v_{mn}^{k}v_{nm}^{j}}{\epsilon_{nm} (\epsilon_{nm}-\hbar\omega)})&&=\frac{e^3}{\omega}\sum_{n\neq m} f_{n}\partial_{i}[ \frac{v_{mn}^{k}v_{nm}^{j}}{\epsilon_{nm} (\epsilon_{nm}-\hbar\omega)}- \frac{v_{nm}^{k}v_{mn}^{j}}{\epsilon_{nm} (\epsilon_{nm}+\hbar\omega)}],\nonumber\\
&&=\frac{e^3}{\omega}\sum_{n\neq m} f_{n}\partial_{i}[\frac{v_{mn}^{k}v_{nm}^{j}-v_{nm}^{k}v_{mn}^{j}}{\epsilon_{nm}^2}]-\hbar e^3\sum_{n\neq m} f_n\partial_{i} \frac{ v_{mn}^{k}v_{nm}^{j}+v_{nm}^{k}v_{mn}^{j}}{\epsilon_{nm}^3}+O(\omega).\nonumber\\
\end{eqnarray}
We have used the small $\omega$ approximation for the second term. The band Berry curvature and normalized quantum metric are defined as
\begin{equation}
\Omega^{i'}_{n}=\sum_{m\neq n} \delta_{i'jk}\frac{v_{mn}^{k}v_{nm}^{j}-v_{nm}^{k}v_{mn}^{j}}{\epsilon_{nm}^2},\ G^{jk}_{n}=\sum_{m\neq n} \frac{ v_{mn}^{k}v_{nm}^{j}+v_{nm}^{k}v_{mn}^{j}}{\epsilon_{nm}^3}.
\end{equation}

Then we find the interband contributions of Fermi surface photocurrents are written as
\begin{equation}
j^{i}_{fs}\approx \frac{e^3}{\omega}\sum_{n} f_{n} \partial_{i} \Omega_{n}^{i'} \delta_{i'jk} E_{j}(\omega)E_{k}(-\omega)-\hbar e^3\sum_n f_{n}\partial_{i} G_{n}^{jk}E_{j}(\omega)E_{k}(-\omega).
\end{equation}
The first term corresponds to the nonlinear Hall term induced by the Berry curvature dipole. The second term represents the intrinsic nonlinear transport induced by quantum metric. 
 \end{document}